\documentclass[10pt,fleqn]{article}

\usepackage{latexsym}
\usepackage{amsbsy}
\usepackage{amsfonts}
\usepackage{amssymb}
\usepackage{amscd}
\usepackage{t1enc}
\usepackage{epsfig}
\usepackage{color}
\usepackage{cite}

\usepackage{caption}

\mathchardef\Gamma="0100
\mathchardef\Delta="0101
\mathchardef\Theta="0102
\mathchardef\Lambda="0103
\mathchardef\Xi="0104
\mathchardef\Pi="0105
\mathchardef\Sigma="0106
\mathchardef\Upsilon="0107
\mathchardef\Phi="0108
\mathchardef\Psi="0109
\mathchardef\Omega="010A

\def\pocha#1{\,\,\raise -.08ex\hbox{\mbox{\LARGE$\check{\,}$}}\hbox{\mbox{$\!\!\!#1$}}}
\def\pochaf#1{\,\,\raise -.5ex\hbox{\mbox{\huge $\check{\,}$}}\hbox{\mbox{$\!\!\!\!#1$}}}%POCHA DECALE POUR LA LETTRE f
\def\pochageant#1{\,\,\raise -1ex\hbox{\mbox{\Huge$\check{\,}$}}\hbox{\mbox{$\!\!\!\!#1$}}}
\def\pochabas#1{\,\,\raise -.5ex\hbox{\mbox{\LARGE$\check{\,}$}}\hbox{\mbox{$\!\!\!#1$}}}
\def\I{{\rm i}}
\def\D{{\rm d}}
\def\E{{\rm e}}

\def\vec{\boldsymbol}

\def\cal{\mathcal}

\input{diagxy}

\begin{document}

\title{A Light-Ray Approach to Fractional Fourier Optics}

\date{}

\maketitle

\begin{center}

\vskip -1.2cm 

{
\renewcommand{\thefootnote}{}
 {\bf   \'Eric Fogret and Pierre Pellat-Finet\footnote{\hskip -.53cm Laboratoire de Math\'ematiques de Bretagne Atlantique, UMR CNRS 6205,

\noindent Universit\'e de Bretagne Sud, CS 60573, 56017 Vannes, France.

\noindent eric.fogret@univ-ubs.fr, pierre.pellat-finet@univ-ubs.fr     %\hfill \today
 }}
}
\setcounter{footnote}{0}

\medskip
{\sl \small Univ Bretagne Sud,   CNRS UMR 6205, LMBA, F-56000 Vannes, France}
 
\end{center}

\vskip .5cm
%*************************************ABSTRACT

\begin{center}
\begin{minipage}{12cm}
\hrulefill

\smallskip
{\small
  {\bf Abstract.}
A light ray in space is characterized by two vectors: (i) a transverse spatial-vector associated with the point where the ray intersects a given spherical cap; (ii) an angular-frequency vector which defines the ray direction of propagation. 
 Given a light ray propagating from a spherical emitter to a spherical receiver, a linear equation is established that links its representative vectors on the emitter and on the receiver. The link is expressed by means of a  matrix which is not homogeneous, since it involves both spatial and angular variables (having distinct physical dimensions). Indeed, the matrix becomes a homogeneous rotation-matrix  after  scaling the previous variables  with appropriate dimensional coefficients. When applied to diffraction, in the framework of a scalar theory, the scaling operation results  directly in  introducing  fractional-order Fourier transformations as mathematical expressions of Fresnel diffraction phenomena. Linking angular-frequency vectors  and spatial frequencies  results in an interpretation of the notion of a spherical angular-spectrum.  Accordance of both inhomogeneous and homogeneous ray-matrices with  Huygens-Fresnel principle is examined. The proposed ray-matrix representation of diffraction is also applied to coherent imaging through a lens.


\smallskip
\noindent {\bf Content}

\smallskip

\noindent 1. Introduction \dotfill \pageref{sect1}

\noindent 2. Space and angular variables and their transfers \dotfill \pageref{sect2}

\noindent 3. Rotations in the reduced phase-space\dotfill \pageref{sect3}

\noindent 4. Link with diffraction and fractional Fourier optics\dotfill\pageref{sect4}

\noindent 5. Link with the spherical angular-spectrum \dotfill \pageref{sect5}

\noindent 6. Accordance with the Huygens--Fresnel principle\dotfill \pageref{sect6}

\noindent 7. Coherent imaging\dotfill \pageref{sect7}

\noindent 8. Conclusion\dotfill \pageref{conc}

\noindent Appendix A. Proof of Eq.\ (\ref{eq45})\dotfill \pageref{appenA}

\noindent Appendix B. Proof og Eq.\ (\ref{eq52})\dotfill \pageref{appenAbis}

\noindent Appendix C. An alternative proof of the conjugation of curvature centers\dotfill \pageref{appenB}

\noindent Appendix D. Checking $a_{21}=0$\dotfill \pageref{appenC}

\noindent Appendix E. An alternative proof of the radius magnification law\dotfill \pageref{appenD}

\noindent Appendix F. Homogeneous imaging-matrix\dotfill \pageref{appenE}

\noindent  References\dotfill \pageref{refer}

}
\hrulefill
\end{minipage}
\end{center}

%**********************************************************************
\section{Introduction}\label{sect1}
%**********************************************************************

The link between fractional-order Fourier transformations and Fresnel diffraction has been the subject of many articles since 1993 \cite{Alie1,PPF1,PPF2,PPF3}. More generally, fractional-order Fourier transformations have been associated with various propagation issues in optics \cite{Men1,Oza2,Oza3}. Those works fall into a subclass of Fourier optics, which we call fractional Fourier optics \cite{PPF3,Oza2}. Among the various methods of fractional Fourier optics, some use matrix representations of light propagation ({\sc abcd}  matrices) for diffraction as well as for imaging through lenses \cite{Ber}. They deal with ray vectors and square matrices, which are inhomogeneous matrices in the sense that they involve spatial variables as well as angular ones (matrix elements have distinct physical dimensions). 

A way of introducing fractional-order Fourier transformations in optics is through Wigner distributions associated with optical fields, which are  phase--space representations including both field amplitudes and their spectra (Fourier transforms) \cite{Loh1}. In such a case, the temptation is high to describe the effect of propagation or imaging as a geometrical isometry, e.g.\ a rotation. It should be clear that this may be done only on a homogeneous space, that is, after having defined an appropriate dimensional scaling of spatial and angular variables, so that reduced variables are dimensionless or have a common physical dimension \cite{PPF4}.

In the present article we use ray vectors and matrices and look for conditions to transform them into homogeneous vectors and matrices. Since we are trying to represent diffraction phenomena, we shall consider coherent fields, and  according to a scalar theory of diffraction, quadratic-phase factors have to be taken into account; the ray-matrix method we introduce is adapted to spherical emitters and receivers, a way of managing with those quadratic-phase factors \cite{PPF3}.

We shall show that looking for transfer ray-matrices being  rotation matrices directly leads us to represent Fresnel diffraction phenomena by  fractional-order Fourier transformations, in accordance with previous works \cite{PPF3}.  We shall then interpret how the proposed ray-matrix method is in accordance with the Huygens-Fresnel principle.

The notion of a spherical angular-spectrum  \cite{SAS1,SAS2} has been introduced as a generalization of the usual ``planar'' angular-spectrum \cite{Goo}. We shall show that a spherical angular-spectrum can be simply interpreted in terms of the proposed ray-matrix theory. 
 We shall eventually apply ray matrices to geometrical coherent imaging.

%*********************************************************************
\section{Space and angular variables and their transfers}\label{sect2}
%*********************************************************************

According to a scalar theory of diffraction, the transfer of the optical-field amplitude from a usually plane emitter to a receiver at a given distance involves quadratic phase factors \cite{PPF3,Goo}. These factors can be handled by using spherical emitters or receivers, that is, spherical caps on which  field amplitudes are defined \cite{PPF3,GB1,GB2}. We begin by adapting to spherical caps a light-ray representation that is currently used in paraxial geometrical optics,  in which a light ray is represented by  a transverse vector and by the angle made by the ray with the optical axis. Since emitters and receivers are spherical caps, calculi will be developed up to second order in function of transverse and angular variables.

\subsection{Angular frequency and light-ray representation}%*****************

\begin{figure}%[b]%$$$$$$$$$$$$$$$$$$$$$$$$$$$$$$$$$$$$$$$$$$$$$$$$$$$$$$$$$$$
  \begin{center}
    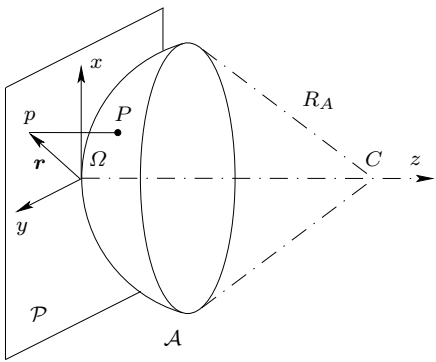
  \end{center}
   \caption{\small Coordinates on a spherical cap ${\cal A}$. A point $P$ on ${\cal A}$ is defined by the coordinates $x$, $y$ of $p$ on the plane ${\cal P}$, tangent to ${\cal A}$ at its vertex $\Omega$. \label{fig1}}
 \end{figure}

Let ${\cal A}$ be a spherical cap, whose vertex is $\Omega$ and center of curvature is $C$ (Fig.\ \ref{fig1}). The radius of curvature of ${\cal A}$ is 
$R_A=\overline{\Omega C}$ (an algebraic measure). Let $P$ be a point on ${\cal A}$ and let $p$ be
the orthogonal projection of $P$ on the plane ${\cal P}$ tangent to ${\cal A}$
at $\Omega$.  We choose Cartesian coordinates $x,y$ on ${\cal P}$, so that $p$ is
perfectly defined by the two-dimensional vector $\vec{Sp}=\vec r=(x_p,y_p)$.

 Numbers  $x_p$  and $y_p$ are the coordinates of $p$, and we say that $\vec r$ is the spatial variable associated with $p$. Given ${\cal A}$, coordinates $x_p$ and $y_p$ can also be used as coordinates of $P$ on the sphere; in the following, indices will be dropped. (The former analysis holds true because the spherical cap ${\cal A}$ is less than half a sphere. In fact in the following, ${\cal A}$ will be close enough to  ${\cal P}$ so that second-order approximations are legitimate.)
 
\begin{figure}[b]%$$$$$$$$$$$$$$$$$$$$$$$$$$$$$$$$$$$$$$$$$$$$$$$$$$$$$$$$$$$$$$$$
\begin{center}
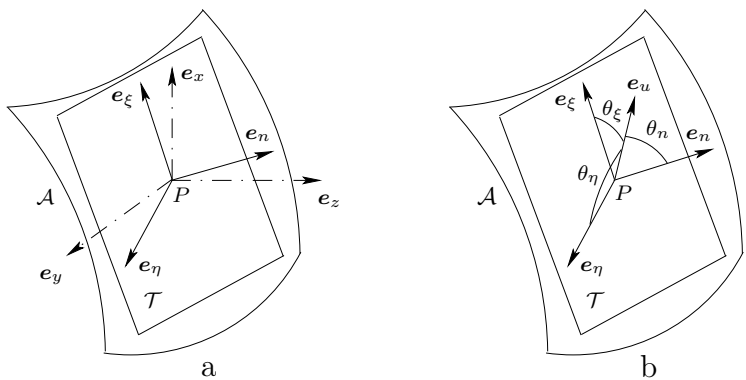
\end{center}
%\vskip -.4cm
\caption{\small (a) Definition of a direct basis at point $P$; (b) Direction cosines of a unit vector $\vec e_u$. \label{fig1quater}}
\end{figure}

Let $z$ be the axis along light propagation and let $\vec e_x$, $\vec e_y$ and $\vec e_z$ be unit vectors along $x$, $y$ and $z$, forming a direct basis ($\vec e_x\vec\times \vec e_y=\vec e_z$). Let $\vec e_n$ be the unit vector, normal to ${\cal A}$ at $P$, so that the Euclidean scalar product $\vec e_n\vec \cdot\vec e_z$ is positive. We introduce the following unit vectors
\begin{equation}
\vec e_\xi={\vec e_y\vec\times \vec e_n\over ||\vec e_y\vec\times \vec e_n ||}\,,\hskip 1cm
\vec e_\eta={\vec e_n\vec\times \vec e_x\over ||\vec e_n\vec\times \vec e_x ||}\,,\end{equation}
such that $\vec e_\xi, \vec e_\eta,\vec e_n$ form a direct basis ($\vec e_\xi \vec\times \vec e_\eta = \vec e_n$). We remark that $\vec e_\xi$ and $\vec e_\eta$ lie in the plane ${\cal T}$, tangent to ${\cal A}$ at $P$ (Fig.\ \ref{fig1quater} a).

%$$$$$$$$$$$$$$$$$$$$$$$$$$$$$$$$$$$$$$$$$$$$$$$$$$$$$$$$$$$$$$$$$

Let us consider a light ray passing through $P$ and let $\vec e_u$ be the unit vector along the direction of pro\-pa\-ga\-tion of the ray (Fig.\ \ref{fig1quater} b). We define the direction cosines of $\vec e_u$ with respect to $\vec e_\xi $, $\vec e_\eta$ and $\vec e_n$ by
\begin{equation}
 \xi =\cos \theta_\xi\,,\hskip .5cm
\eta =\cos \theta_\eta\,,\hskip .5cm
\zeta =\cos \theta_n\,,\end{equation}
where $\theta_\xi$ is the angle between $\vec e_\xi$ and $\vec e_u$, etc.\  (Fig.\ \ref{fig1quater} b). Since $\vec e_u$ is a unit vector, we have $\xi^2+\eta^2+\zeta^2=1$, so that $\vec e_u$ is perfectly defined by $\xi$ and $\eta$ (because we impose $\zeta >0$, for waves propagating along positive $z$).

%$$$$$$$$$$$$$$$$$$$$$$$$$$$$$$$$$$$$$$$$$$$$$$$$$$$$$$$$$$$$$$$$$$$

\begin{figure}%[b]%$$$$$$$$$$$$$$$$$$$$$$$$$$$$$$$$$$$$$$$$$$$$$$$$$$$$$$$$$$$$$$$$
  \begin{center}
    %\vskip -.6cm
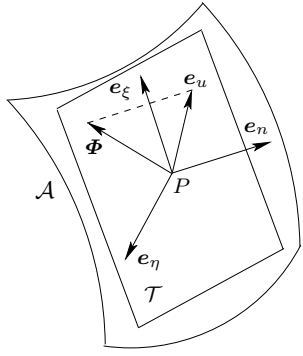
\end{center}
%\vskip -.4cm
\caption{\small Given a ray along the unit vector $\vec e_u$, the corresponding angular-frequency vector $\Phi$ is the projection of $\vec e_u$ on the plane tangent to ${\cal A}$ at $P$. \label{fig2}}
\end{figure}

We call angular-frequency vector the two-dimensional vector $\vec \Phi$ defined by
\begin{equation}
\vec \Phi =(\xi , \eta )\,.\end{equation}
The vector $\vec \Phi$ is the projection of $\vec e_u$ on the plane tangent to ${\cal A}$ at $P$ (Fig.\ \ref{fig2}). It is an element of ${\mathbb R}^2$ and has no physical dimension.

A light ray coming from ${\cal A}$ is perfectly defined by the ordered pair $(\vec r
,\vec \Phi )$, so that we shall say ``the ray $(\vec r
,\vec \Phi )$''.

%$$$$$$$$$$$$$$$$$$$$$$$$$$$$$$$$$$$$$$$$$$$$$$$$$$$$$$$$$$$$$$$$$$$

\subsection{Ray transfer}%****************************

Let ${\cal B}$ be a spherical receiver at a distance $D$ from ${\cal A}$ ($D=\overline{\Omega\Omega '}$, where $\Omega '$ is the vertex of ${\cal B}$). The radius of curvature of ${\cal B}$ is $R_B$. A ray $(\vec r,\vec \Phi )$ issued from $P$ on  ${\cal A}$ intersects ${\cal B}$ at $P'$, where the ray is defined  by $(\vec r',\vec \Phi ')$; our first task is to find the link between $(\vec r,\vec \Phi )$ and $(\vec r',\vec \Phi ')$, that is, between $(x,y,\xi , \eta )$ and $(x',y',\xi',\eta ')$.

We are looking for relations of the form
$x'=x'(x,y,\xi,\eta )$ ($x'$ is a function of $x$, $y$, $\xi$ and $\eta$),  $\xi '=\xi '(x,y,\xi,\eta )$ etc.\ 
and restrict ourselves to second-order approximations with respect to transverse variables: we  neglect terms whose orders are greater than or equal to 3.
For example $x'$ is written as
\begin{eqnarray}
x'&\!\!\!=&\!\!\!a_0+a_1x+b_1y+c_1\xi +d_1\eta +a_2x^2+b_2y^2+e_2xy+c_2\xi^2+d_2\eta^2 \nonumber \\
& & \hskip 6cm +f_2\xi\eta +g_2x\xi +h_2x\eta+k_2y\xi +m_2y\eta\,.\label{eq9}\end{eqnarray}

\begin{figure}[h]%$$$$$$$$$$$$$$$$$$$$$$$$$$$$$$$$$$$$$$$$$$$$$$$$$$$$$$$$$$$$$$
\begin{center}
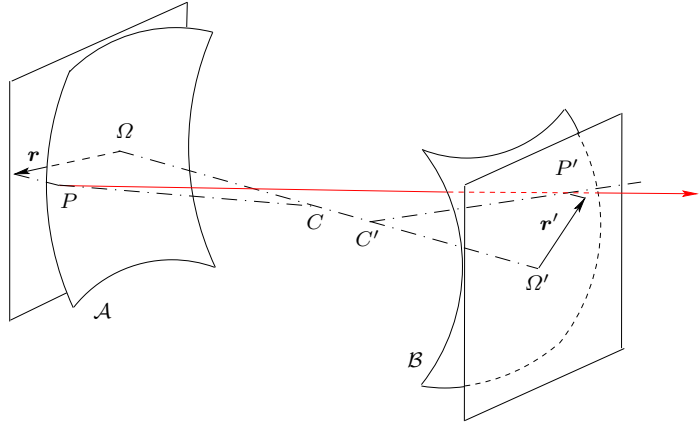
\end{center}
\caption{\small Elements for ray transfer from $P$ (on the emitter ${\cal A}$) to $P'$ (on the receiver ${\cal B}$).\label{fig3}}
\end{figure}

Now we remark that if we rotate ${\cal A}$ and ${\cal B}$ (Fig.\ \ref{fig3}) by
an angle $\pi$ around the $z$ axis, we change $\vec r$ into its opposite
$-\vec r$, and the same for $\vec \Phi$, $\vec r'$ and $\vec \Phi '$. Thus
$x$, $y$, $\xi$ and $\eta$ are changed into their opposites, as well as
$x'$. Consequently $a_0$ and the second-order terms in Eq.\ (\ref{eq9}) must vanish so that, within a second-order approximation, $x'$ is a linear function of $x$, $y$, $\xi$ and $\eta$, that is,  
\begin{equation}
  x'=a_1x+b_1y+c_1\xi +d_1\eta\,.\end{equation}
%$$$$$$$$$$$$$$$$$$$$$$$$$$$$$$$$$$$$$$$$$$$$$$$$$$$$$$$$$$$$$$$$$$$

The same can be done with $y'$, $\xi '$ and $\eta '$, so that the relation between $(\vec r,\vec \Phi )$ and $(\vec r',\vec \Phi' )$ is linear. We adopt a matrix form and write
\begin{equation}
  \begin{pmatrix}{x'\cr y'\cr \xi '\cr \eta '}\end{pmatrix}=
  \begin{pmatrix}{a_{11}& a_{12}& a_{13} & a_{14}\cr
      a_{21}& a_{22}& a_{23} & a_{24}\cr
        a_{31}& a_{32}& a_{33} & a_{34}\cr
          a_{41}& a_{42}& a_{43} & a_{44}\cr}\end{pmatrix}
   \begin{pmatrix}{x\cr y\cr \xi \cr \eta }\end{pmatrix}\,. \label{eq8}
  \end{equation}

\begin{figure}[b]%$$$$$$$$$$$$$$$$$$$$$$$$$$$$$$$$$$$$$$$$$$$$$$$$$$$$$$$$$$$$$$$$$$
\begin{center}
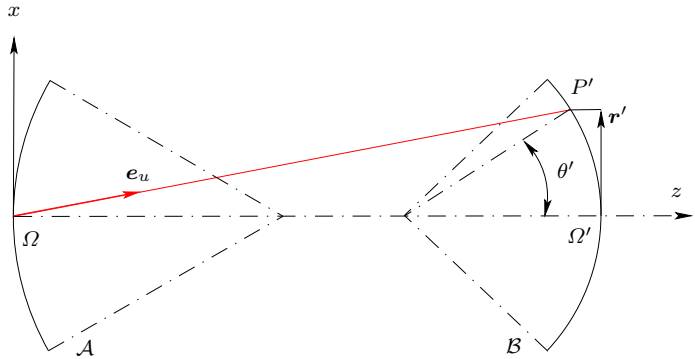
\end{center}
\caption{\small A ray corresponding to $\vec r=0$. The distance from ${\cal A}$ to ${\cal B}$ is taken from vertex to vertex: $D=\overline{\Omega \Omega '}$.\label{fig4}}
\end{figure}%$$$$$$$$$$$$$$$$$$$$$$$$$$$$$$$$$$$$$$$$$$$$$$$$$$$$$$$$$$$$$$$$$$$$$$$$

The coefficients $a_{ij}$ can be determined by examining special cases, as follows.
\begin{itemize}
  \item[i--]
We first assume that $\vec r=0$ (Fig.\ \ref{fig4}). Then $\vec r'=D\vec
\Phi$ (second-order approximation in $\xi$ and $\eta$), that is $(x',y')=(D\xi , D\eta)$,
and we conclude by $a_{13}=a_{24}=D$, and $a_{14}=a_{23}=0$.

\item[ii--]

 We then assume  that $\vec e_u$ is in the $x$--$z$ plane  (Fig.\ \ref{fig4}): $\vec
e_u=u_x\vec e_x + u_z\vec
e_z$, with ${u_x}^2+{u_z}^2=1$. For $\vec r=0$, we have $(\vec
e_\xi ,\vec e_\eta ,\vec e_n)= (\vec
e_x ,\vec e_y ,\vec e_z)$, so that
\begin{equation}
\vec \Phi =(u_x ,0)\,.\end{equation}
We introduce the angle $\theta '$ (Fig.\ \ref{fig4}) and we obtain
\begin{equation}
\vec e_{\xi '}=\vec e_x\cos\theta '-\vec e_z\sin\theta '\,, \hskip .5cm 
\vec e_{\eta '}=\vec e_y\,, \hskip .5cm \vec e_n=\vec e_x\sin\theta '+\vec
e_z\cos\theta ',\end{equation}
and
\begin{equation}
\vec \Phi' =(u_x \cos\theta '-u_z\sin\theta ' , 0)\,.\end{equation}
In the limits of a second-order approximation, we have
\begin{equation}
\sin\theta '=-{Du_x \over R_B}\,,\;\hskip 1cm \cos\theta ' =1-{D^2{u_x}^2\over
  2{R_B}^2}\,,\end{equation}
and then
\begin{equation}
u_x \cos\theta '-u_z\sin\theta '={D+R_B\over R_B}\,u_x\,,\end{equation}
so that 
\begin{equation}
(\xi ',0)=\vec \Phi '={D+R_B\over R_B}\, \vec \Phi ={D+R_B\over R_B} (\xi ,0)\,.\end{equation}
We conclude by $a_{34}=0$, and 
\begin{equation}
  a_{33}={D+R_B\over R_B}\,.\end{equation}

The same reasonning in the $y$--$z$ plane leads to $a_{43}=0$, and
\begin{equation}
  a_{44}={D+R_B\over R_B}\,.\end{equation}

\item[iii--]
 
We now assume  $\vec r\ne 0$ and $\vec \Phi =0$. Figure \ref{fig5} shows that
\begin{equation}
(x',y')=\vec r'={R_A-D\over R_A}\vec r= {R_A-D\over R_A} (x,y)\,,\end{equation}
and we conclude by $a_{12}=a_{21}=0$, and
\begin{equation}
  a_{11}=a_{22}={R_A-D\over R_A}\,.\end{equation}

\item[iv--]
Finally, by choosing $\vec e_u$ in the $x$--$z$ plane, we have $\vec u =u_x \vec e_x
+u_z\vec e_z$ and we introduce $\theta '$ (Fig. \ref{fig5}) so that
\begin{equation}
\sin\theta '=-{r'\over R_B}\,,\hskip 1cm \cos\theta '=1-{r'^2\over
  2{R_B}^2}\,.\end{equation}

\begin{figure}[b]%$$$$$$$$$$$$$$$$$$$$$$$$$$$$$$$$$$$$$$$$$$$$$$$$$$$$$$$$
\begin{center}
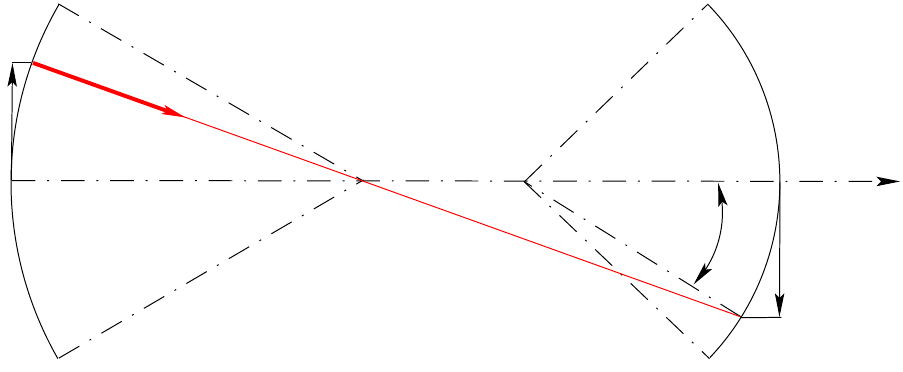
\end{center}
\caption{\small A ray corresponding to $\vec \Phi =0$: the unit vector $\vec e_u$ is orthogonal to ${\cal A}$. \label{fig5}}
\end{figure}%$$$$$$$$$$$$$$$$$$$$$$$$$$$$$$$$$$$$$$$$$$$$$$$$$$$$$$$$$$$$$$
We have
\begin{equation}
\vec \Phi' =(u_x \cos\theta '-u_z\sin\theta ' , 0)\,.\end{equation}
We also have $u_x ={r/ R_A}$,
so that
\begin{equation}
u_x \cos\theta '-u_z\sin\theta '={r\over R_A}+{R_A-D\over R_B}r
=-{D-R_A+R_B\over R_AR_B}r\,,
\end{equation}
and
\begin{equation}
(\xi ',0)=\vec \Phi '=-{D-R_A+R_B\over R_AR_B}\vec r={R_A-R_B-D\over R_AR_B} (x,0) \,.\end{equation}
We conclude by $a_{32}=0$, and
\begin{equation}
  a_{31}={R_A-R_B-D\over R_AR_B}\,.\end{equation}

The same reasoning in the $y$--$z$ leads to $a_{41}=0$, and
\begin{equation}
  a_{42}={R_A-R_B-D\over R_AR_B}\,.\end{equation}
All the $a_{ij}$'s have been determined.
\end{itemize}

\medskip
In conclusion, Eq.\ (\ref{eq8}) is explicitly written
\begin{equation}
\begin{pmatrix}{ x' \cr y' \cr \xi ' \cr \eta ' }\end{pmatrix}
=\begin{pmatrix}{ \displaystyle{R_A-D \over R_A} & 0 & D & 0 \cr \cr
  0  &\displaystyle{R_A-D \over R_A} &0 & D \cr \cr\displaystyle{R_A-R_B-D\over R_AR_B}& 0 &\displaystyle{D+R_B\over R_B} & 0  \cr \cr  0 &\displaystyle{R_A-R_B-D\over R_AR_B}& 0 & \displaystyle{D+R_B\over R_B} }\end{pmatrix}
\begin{pmatrix}{ x \cr y\cr \xi \cr  \eta }\end{pmatrix}   \,. \label{eq25}
\end{equation}
 With $^t(\vec r,\vec \Phi)=\, ^t(x,y,\xi,\eta)$, a more concise form of Eq.\ (\ref{eq25}) is
\begin{equation}
\begin{pmatrix}{\vec r' \cr \vec \Phi ' }\end{pmatrix}=
\begin{pmatrix}{ \displaystyle{R_A-D \over R_A} & D \cr
    \cr
 \displaystyle{R_A-R_B-D\over R_AR_B} &  \displaystyle{D+R_B\over R_B} }\end{pmatrix}
\begin{pmatrix}{ \vec r \cr \vec \Phi  }\end{pmatrix}\,,\label{eq26}
\end{equation}

Eventually we point out that Eqs.\ (\ref{eq25}) and (\ref{eq26}) hold for skew rays as well as for meridional rays.

%**********************************************************
\section{Rotations in a reduced phase-space}\label{sect3}
%**********************************************************

\subsection{Defining reduced variables and an angle of rotation}\label{sect31}%**************

The previous matrices---Eqs.\ (\ref{eq25}) and (\ref{eq26})---are not homogeneous, because $\vec r$ and $\vec r'$ are spatial vectors (their components are homogeneous to lengths) while $\vec \Phi$ and $\vec \Phi '$ are physically dimensionless. We would like to express Eq.\ (\ref{eq26}) as
\begin{equation}\begin{pmatrix}{ \vec \rho' \cr \vec \phi '} \end{pmatrix}=
\begin{pmatrix}{\cos\alpha & \sin\alpha \cr -\sin\alpha & \cos\alpha }\end{pmatrix}\begin{pmatrix}{ \vec \rho \cr \vec \phi  }\end{pmatrix}\,,\label{eq24}
\end{equation}
where $\vec \rho$, $\vec \phi$, $\vec \rho '$ and $\vec \phi '$ are reduced variables that are dimensionless (from a physical point of view) and which replace $\vec r$, $\vec \Phi$, $\vec r$ and $\vec \Phi '$.
(Mathematically, they are two-dimensional vectors, elements of ${\mathbb R}^2$.) 

The square matrix in Eq.\ (\ref{eq24}) is a rotation matrix of angle $-\alpha$. We  choose a rotation angle equal to $-\alpha$ to match $\alpha$ with the forthcoming fractional parameter $\alpha$ associated with a diffraction phenomenon. The value of the fractional parameter ($\alpha$ or $-\alpha$) is related to the definition of two-dimensional fractional-order Fourier transformations, according to Eq.\ (\ref{eq47}).

To express the matrix in Eq.\ (\ref{eq26}) as that in Eq.\ (\ref{eq24}), we set $\vec \rho = A\vec r$, $\vec \rho '=A'\vec r'$, $\vec \phi =B\vec \Phi$ and $\vec \phi '=B'\vec \Phi$, where $A$, $A'$, $B$ and $B'$ are positive real numbers. We should have then
\begin{equation}
{A\over A'}\cos\alpha ={R_A-D\over R_A}\,,\label{eq28}\end{equation}
\begin{equation}
{A\over B'}\sin\alpha =-{R_A-R_B-D\over R_AR_B}\,,\end{equation}
\begin{equation}
{B\over A'}\sin\alpha =D\,,\label{eq30}\end{equation}
\begin{equation}
  {B\over B'}\cos\alpha ={D+R_B\over R_B}\,.\label{eq31}\end{equation}

Before calculating $A$, $A'$, $B$ and $B'$, we define $\alpha$. Thus we introduce
\begin{equation}
  J={(R_A-D)(D+R_B)\over D(D-R_A+R_B)}\,,\end{equation}
and assume $J\ge 0$,
so that we can choose $\alpha$, with  $-\pi <\alpha <\pi$, such that
\begin{equation}
  \cot^2\alpha ={(R_A-D)(D+R_B)\over D(D-R_A+R_B)}=J\,.\label{eq33}\end{equation}
(If $J<0$, the parameter $\alpha$ becomes a complex number \cite{PPF3,ComplX,PPF6}.)

To complete the definition of $\alpha$, we note that according to Eq.\ (\ref{eq30}), the sign of $\alpha$ should be that of $D$, and according to Eqs.\ (\ref{eq28}) and (\ref{eq30}), the sign of $\cot\alpha$ should be the sign of $(R_A-D)DR_A$. Thus, in addition to Eq.\ (\ref{eq33}) we choose $\alpha\in\; ]-\pi,\pi [$, with $\alpha D\ge 0$ and
      \begin{equation}
        {R_AD\over R_A-D}\,\cot\alpha \ge 0\,.\end{equation}

      On the other hand, we note that according to Eqs.\ (\ref{eq30}) and (\ref{eq31}) the sign of $\cot\alpha$ should also be the sign of $(R_B+D)DR_B$. To avoid inconsistency between Eqs.\ (\ref{eq28}--\ref{eq31}) and the  previous
      definition of $\alpha$, we have to prove that $(R_A-D)DR_A$ and $(R_B+D)DR_B$ have the same sign. Actually, this is a consequence of the assumption $J\ge 0$. For a proof, we start with the identity
      $D(D-R_A-R_B)=R_AR_B-(R_A-D)(D+R_B)$,
      and deduce, for $J\ge 0$,
      \begin{equation}
        {R_AR_B\over (R_A-D)(D+R_B)}=1+{1\over J}\ge 1\,.
      \end{equation}
      We then obtain
       \begin{equation}
        {R_AR_BD^2\over (R_A-D)(D+R_B)}\ge 0\,,
       \end{equation}
       and we conclude that $R_AD(R_A-D)$ and $R_BD(D+R_B)$ have the same sign. The definition of $\alpha$ is consistent with Eqs.\ (\ref{eq28}--\ref{eq31}).

Now, it should be clear that $A$, $A'$, $B$ and $B'$ are defined up to a multiplicative factor. To make the link with diffraction, we shall impose
\begin{equation}
\vec \rho \vec \cdot\vec \phi ={1\over \lambda}\,\vec r\vec \cdot \vec \Phi\,,\label{eq37s}\end{equation}
and
\begin{equation}
\vec \rho '\vec \cdot\vec \phi '={1\over \lambda}\,\vec r'\vec \cdot \vec \Phi '\,.\label{eq38s}\end{equation}
Eventually, from Eqs.\ (\ref{eq28}--\ref{eq31}), we  obtain
\begin{equation}
A^4={1\over \lambda^2{R_A}^2}{(R_A-D)(D-R_A+R_B)\over D(D+R_B)}\,,\label{eq36}\end{equation}
\begin{equation}
B^4={{R_A}^2\over \lambda^2}{D(D+R_B)\over (R_A-D)(D-R_A+R_B) }\,,\label{eq37}\end{equation}
\begin{equation}
A'^4={1\over \lambda^2{R_B}^2}{(D+R_B)(D-R_A+R_B)\over D(R_A-D)}\,,\label{eq38}\end{equation}
\begin{equation}
  B'^4={{R_B}^2\over \lambda^2}{D(R_A-D)\over (D+R_B)(D-R_A+R_B) }\,.\label{eq39}\end{equation}

\subsection{Interpreting rotations in the reduced phase-space}%**********************

We remark that Eq.\ (\ref{eq25}) can also be written
\begin{equation}
\begin{pmatrix}{ x' \cr \xi '\cr y' \cr \eta ' }\end{pmatrix}
=\begin{pmatrix}{ \displaystyle{R_A-D \over R_A} & D & 0 & 0 \cr \cr\displaystyle{R_A-R_B-D\over R_AR_B}&\displaystyle{D+R_B\over R_B} & 0 & 0 \cr \cr 0 & 0 &\displaystyle{R_A-D \over R_A} & D \cr \cr 0 &0 &\displaystyle{R_A-R_B-D\over R_AR_B} & \displaystyle{D+R_B\over R_B} }\end{pmatrix}
\begin{pmatrix}{ x \cr \xi \cr y \cr \eta }\end{pmatrix}   \,. \label{eq40}
\end{equation}
or equivalently, with reduced variables,
\begin{equation}
  \begin{pmatrix}{\rho_x '\cr \phi_x'\cr \rho_y'\cr\phi_y'}\end{pmatrix}=
  \begin{pmatrix}{\cos\alpha & \sin\alpha & 0 & 0\cr
      -\sin\alpha & \cos\alpha & 0 & 0 \cr
      0 &0& \cos\alpha & \sin\alpha \cr
      0 &0& -\sin\alpha & \cos\alpha }\end{pmatrix} 
   \begin{pmatrix}{\rho_x \cr \phi_x\cr \rho_y\cr\phi_y}\end{pmatrix}\,,\label{eq41}
\end{equation}
where $\vec \rho =(\rho_x,\rho_y)$ and $\vec \phi =(\phi_x,\phi_y)$.

Equation (\ref{eq41}) shows that in the reduced phase-space the transfer from ray $(\vec r,\vec \Phi)$ to ray $(\vec r',\vec \Phi ')$ is represented by a 4--dimensional rotation that splits into two  rotations of angle $-\alpha$, each rotation operating in a 2--dimensional subspace of the reduced phase-space \cite{PPF4}. From a physical point of view, matrices in Eq.\ (\ref{eq41}) are dimensionless.

%*************************************************************************
\section{Link with diffraction and fractional Fourier optics}\label{sect4}
%*************************************************************************

\subsection{General transfer by diffraction (Fresnel phenomenon)}

According to a scalar theory of diffraction, the field transfer from ${\cal A}$ to ${\cal B}$ is expressed as \cite{PPF2,PPF3,GB1,GB2}
\begin{eqnarray}
U_B(\vec r')\!\!\!&=&\!\!\!{\I\over \lambda D}\exp \left[-{\I\pi \over \lambda}\left({1\over R_B}+{1\over D}\right)r'^2\right]\nonumber \\
& &\hskip 1.5cm  \times \;\int_{{\mathbb R}^2}
\exp \left[-{\I\pi \over \lambda}\left({1\over D}-{1\over R_A}\right)r^2\right]\,
\exp\left({2\I\pi\over \lambda D}\vec r'\vec\cdot\vec r\right)\,U_A(\vec r)\,\D \vec r\,,\label{eq42}\end{eqnarray}
where $U_A$ is the field amplitude on ${\cal A}$, $U_B$ the field amplitude on ${\cal B}$, and $\D\vec r=\D x\,\D y$. (A phase factor equal to $\exp (-2\I\pi D/\lambda )$ has been omitted.)

We replace $\vec r$ and $\vec r'$ by reduced variables $\vec \rho$ and $\vec \rho '$, as defined in Sect.\ \ref{sect31}, and we use reduced field amplitudes defined by
\begin{equation}
V_A(\vec \rho )=\sqrt{\left|R_AD\over R_A-D\right|}\;U_A\left({\vec \rho\over A}\right)\label{eq43}\,,\end{equation}
\begin{equation}
V_B(\vec \rho ')=\sqrt{\left|R_BD\over R_B+D\right|}\;U_B\left({\vec \rho '\over A'}\right)\,.\label{eq44}\end{equation}
Then Eq.\ (\ref{eq42}) becomes (the proof is given in Appendix A)
\begin{equation}
V_B(\vec \rho')={\I\over \sin\alpha }\exp (-\I\pi\rho'^2\cot\alpha)\int_{{\mathbb R}^2}
\exp (-\I\pi \rho^2\cot\alpha)\,
\exp\left({2\I\pi\over \sin\alpha }\,\vec \rho '\vec\cdot\vec \rho\right)\,V_A(\vec \rho)\,\D \vec \rho\,,\label{eq45}\end{equation}
that is
\begin{equation}
  V_B(\vec \rho ')=\E^{\I\alpha}\,{\cal F}_{\alpha} [V_A](\vec \rho ')\,,\label{eq46}\end{equation}
where ${\cal F}_{\alpha }$ denotes the fractional Fourier transformation of order $\alpha$, defined, for a two-dimensional function $f$, by \cite{Nam,Mcb}
\begin{equation}
{\cal F}_\alpha [f](\vec \rho )={\I \E^{-\I\alpha}\over\sin\alpha}\,\exp (-\I\pi\rho'^2\cot\alpha)\,\int_{{\mathbb R}^2}
\exp (-\I\pi \rho^2\cot\alpha)\,
\exp\left({2\I\pi\over \sin\alpha }\,\vec \rho '\vec\cdot\vec \rho\right)\,f(\vec \rho)\,\D \vec \rho\,.
\label{eq47}
\end{equation}

Equation (\ref{eq46}) is usually deduced in the framework of fractional Fourier optics \cite{PPF2,PPF3}, by choosing appropriate reduced variables and reduced field amplitudes. Reduced variables have been introduced here with the help of ray-matrices by looking for homogeneous matrices. In other words, the basic equation that expresses diffraction in the framework of fractional Fourier optics has been established from the analysis of  ray transfers from an emitter to a receiver and  considering homegenous ray-matrices.

Equation (\ref{eq42}) generally corresponds to a Fresnel-diffraction phenomenon \cite{PPF1,PPF2,PPF3}. Fraunhofer diffraction constitutes a special case and is the subject of the next section.

\subsection{Fraunhofer diffraction}

\begin{figure}[b]%$$$$$$$$$$$$$$$$$$$$$$$$$$$$$$$$$$$$$$$$
\begin{center}
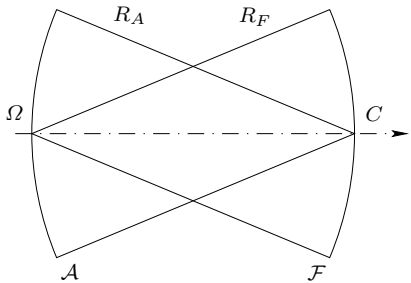
\end{center}
\caption{\small Fraunhofer diffraction. The spherical cap ${\cal F}$ is the Fourier sphere of ${\cal A}$: its radius $R_F$ is $R_F=\overline{C\Omega}=-R_A$.\label{fig6}}
\end{figure}%$$$$$$$$$$$$$$$$$$$$$$$$$$$$$$$$$$$$$$$$$$$$$$

Fraunhofer diffraction \cite{PPF3} occurs when $R_A=D=-R_B$ (Fig.\ \ref{fig6}). Then Eq.\ (\ref{eq42}) takes the form
\begin{equation}
  U_B(\vec r')={\I\over\lambda D}\,\int_{{\mathbb R}^2}\exp\left({2\I\pi\over\lambda D}\vec r'\vec\cdot\vec r\right)\,U_A(\vec r)\,\D\vec r\,,
\end{equation}
and involves a (standard) Fourier transform, that is
\begin{equation}
  U_B(\vec r')={\I\over\lambda D}\,\widehat{U}_A\left({\vec r '\over\lambda D}\right)\,.\end{equation}
The spherical cap ${\cal B}$ is called the Fourier sphere of ${\cal A}$ and will be denoted ${\cal F}$. Spherical caps ${\cal A}$ and ${\cal F}$ are (symmetrical) confocal spheres: the vertex of the one is the curvature center of the other ($R_F=-R_A$). We write
\begin{equation}
  U_F(\vec r')={\I\over\lambda D}\,\widehat{U}_A\left({\vec r '\over\lambda D}\right)={\I\over \lambda R_A}\,\widehat U_A\left({\vec r '\over\lambda R_A}\right)\,.\label{eq50}\end{equation}
If $\tilde f$ denotes the function defined by $\tilde f(\vec r)=f(-\vec r)$, from Eq.\ (\ref{eq50}) we deduce ($\vec F$ is a spatial frequency)
\begin{equation}
  \widehat U_F(\vec F)={\I\lambda R_A}\,\widehat{\!\widehat U}_A(\lambda R_A\vec F)=\I\lambda R_A \widetilde U_A(-\lambda R_A\vec F)\,,\end{equation}
and then
\begin{equation}
  U_A(\vec r)={\I\over\lambda R_F}\,\widehat{U}_F\left({\vec r\over\lambda R_F}\right)\,,\end{equation}
which shows that ${\cal A}$ is the Fourier sphere of ${\cal F}$ (reciprocity property of Fourier spheres).

According to Eq.\ (\ref{eq33}) we have $\cot\alpha =0$, and for positive $D$ we obtain $\alpha =\pi /2$:  Equation (\ref{eq46}) involves a (standard) Fourier transformation, as expected. To deal with reduced variables, we proceed as follows. We consider first that $R_A=-R_F\ne -D$  ($R_F=R_B$), so that according to Eqs.\ (\ref{eq36}) and (\ref{eq38}) we have
\begin{equation}
  A^4={2R_A-D\over \lambda^2{R_A}^2D}\,,\hskip 1cm  A'^4={2R_A-D\over \lambda^2{R_F}^2D}\,.
  \end{equation}
When $D$ tends to $R_A$ we obtain then that $\vec \rho$ tends to $\vec r/\sqrt{\lambda R_A}$ and $\vec\rho '$ tends to  $\vec r'/\sqrt{\lambda R_A}$. Hence reduced variables are perfectly defined.

Thus, for $D>0$, Fraunhofer diffraction is decribed by a rotation of angle $-\alpha =-\pi /2$.  
If $D<0$, we obtain $\alpha =-\pi /2$; we have thus a virtual  Fraunhofer diffraction, described by a rotation of angle $-\alpha =\pi /2$.

%**************************************************************
\section{Link with the spherical angular-spectrum}\label{sect5}
%**************************************************************

\subsection{The notion of spherical angular-spectrum}
The notion of a spherical angular-spectrum is a generalization of the planar angular spectrum to spherical caps \cite{SAS1,SAS2}. We shall provide  an interpretation of the spherical angular-spectrum by linking it with the previous analysis

We begin by associating a light ray and a point on an emitter with a spatial frequency. Let ${\cal A}$ be a spherical emitter (or receiver) on which the field amplitude is
\begin{equation}
U_A(\vec r)=U_0\,\exp (-2\I\pi \vec F_0\vec \cdot \vec r)\,,\label{eq48}\end{equation}
where $U_0$ is a dimensional constant. The vector $\vec F_0$ is a spatial frequency. 

\begin{figure}[b]%********************************************************************
\begin{center}
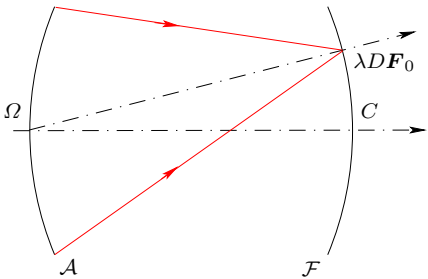
\end{center}
\caption{\small If the field amplitude on ${\cal A}$ takes the form $\exp (-2\I\pi \vec F_0\vec \cdot \vec r)$, the wave issued from ${\cal A}$ converges at point $\vec r'=\lambda D\vec F_0$ on ${\cal F}$ (the Fourier sphere of ${\cal A}$), where $D=R_A=\overline{\Omega C}$.\label{fig7}}
\end{figure}

According to Eq.\ (\ref{eq50}) the field amplitude on ${\cal F}$ (the Fourier sphere of ${\cal A}$) is
\begin{equation}
  U_F(\vec r')={\I\over \lambda D}\,\delta \left({\vec r'\over \lambda D}-F_0\right)=\I\lambda D\,\delta (\vec r'-\lambda D \vec F_0)\,,\end{equation}
where $\delta$ denotes the (2--dimensional) Dirac distribution. We conclude that  the wave emitted by ${\cal A}$ converges at the point $P'$ of ${\cal F}$, such that $\vec r'=\lambda D \vec F_0$ (Fig.\ \ref{fig7}).

Let $P$ be a point on ${\cal A}$. The ray $PP'$ is defined by $(\vec r ,\vec
\Phi_0)$. Since $D=R_A$ (Fourier sphere), Eq.\ (\ref{eq26}) gives $P'$ as  corresponding to 
$\vec r'=D\vec \Phi _0$, so that
\begin{equation}
\vec \Phi_0=\lambda \vec F_0\,.\label{eq51}\end{equation}

%**************************************************************************

Equation (\ref{eq51}) leads us to introduce the notion of a spherical angular-spectrum as follows. Let $U_A$ be the field amplitude on ${\cal A}$. The Fourier transform of $U_A$ (also called the spectrum of $U_A$) is $\widehat U_A(\vec F)$. By changing $\vec F$ into $\vec \Phi =\lambda \vec F$, we obtain the so-called spherical angular-spectrum of $U_A$, denoted $S_A$ and such that
\begin{equation}
S_A(\vec \Phi )= {1\over \lambda^2}\,\widehat U_A\left({\vec \Phi \over \lambda }\right)
={1\over \lambda^2}\int_{{\mathbb R}^2}\exp \left({2\I\pi \over \lambda}\vec \Phi\vec \cdot\vec r\right)\,U_A(\vec r)\,\D \,\vec r\,,\end{equation}
where the factor $1/\lambda^2$ has been introduced for the sake of homogeneity (that is, $S_A$ has the (physical) dimension of $U_A$).

If ${\cal A}$ becomes a plane, $S_A$ is the usual angular-spectrum  (up to a factor $1/\lambda^2$) \cite{Goo}. According to classical Fourier optics, a plane wave is associated with  each spatial frequency of the wave emitted by a planar object. The emitted wave is decomposed on a family of plane waves. Each plane wave propagates along a direction whose direction cosines are given by $(\cos\theta_x,\cos\theta_y)=\vec \Phi =\lambda \vec F$. The vector $\vec \Phi$ is a constant.

The spherical angular-spectrum is defined on a spherical emitter (or receiver). A spherical wave is associated with each spatial frequency on the spherical emitter, so that the emitted wave is decomposed on a family of spherical waves. Every spherical wave is weighted by an appropriate coefficient which is equal to the value of the angular spectrum for the associated spatial frequency.
The law $\vec \Phi =\lambda \vec F$ still holds for spherical emitters (or receivers). But on the basis of the analysis of Sect.\ \ref{sect2}, given a ray propagating along the unit vector $\vec e_u$ and issued from a given  point on an emitter, the angular frequency $\vec \Phi$ should be interpreted as the projection of $\vec e_u$ on the plane tangent to the emitter at the previous point.

\subsection{Propagation of the spherical angular-spectrum}

The transfer of the spherical angular-spectrum from an emitter ${\cal A}$ (radius $R_A$) to a receiver ${\cal B}$ (radius $R_B$) at a distance $D$ is given by \cite{SAS1,SAS2}
\goodbreak
\begin{eqnarray}
  S_B(\vec \Phi ')&\!\!\!\!=&\!\!\!\!{\I R_AR_B\over \lambda (D-R_A+R_B)}\,\exp\left({-\I\pi R_B(R_A-D)\over \lambda (D-R_A+R_B)}\Phi '^2 \right) \label{eq61s}\\
  & & \times\int_{{\mathbb R}^2}\exp\left({-\I\pi R_A(R_B+D)\over \lambda (D-R_A+R_B)}\Phi ^2 \right)
  \exp\left({2\I\pi R_AR_B\over  \lambda (D-R_A+R_B)}\,\vec \Phi '\vec\cdot\vec\Phi \right)\,S_A(\vec \Phi)\,\D\vec \Phi\,.
  \nonumber
\end{eqnarray}
If we make the following changes
\begin{equation}
  D\;\,\;\longmapsto\;{D-R_A+R_B\over R_AR_B}\,,
  \end{equation}
\begin{equation}
  R_A\;\longmapsto\;{D-R_A+R_B\over R_AD}\,,
\end{equation}
\begin{equation}
  R_B\;\longmapsto\;{D-R_A+R_B\over R_BD}\,,
  \end{equation}
in Eq.\ (\ref{eq42}) and replace $\vec r$ by $\vec \Phi$ and $\vec r'$ by $\vec \Phi '$, we obtain Eq.\ (\ref{eq61s}).

We use reduced (vectorial) variables $\vec \phi$ and $\vec \phi '$, as defined in Sect.\ \ref{sect31}, and reduced spherical angular-spectra $T_A$ and $T_B$ defined by
\begin{equation}
  T_A (\vec \phi )=\sqrt{\left|{R_A-D\over R_AD}\right|}\;S_A\left({\vec \phi\over B}\right)\,, %=B^2\,\widehat V_A(\vec \phi )\,,
  \label{eq63s}
\end{equation}
\begin{equation}
  T_B (\vec \phi ')=\sqrt{\left|{R_B+D\over R_BD}\right|}\;S_B\left({\vec \phi '\over B'}\right)\,, %=B'^2\,\widehat V_B(\vec \phi ')\,,
  \label{eq64s}
\end{equation}
so that the propagation of the spherical angular-spectrum from ${\cal A}$ to ${\cal B}$ is expressed as
\begin{equation}
  T_B(\vec \phi ')=\E^{\I\alpha}\,{\cal F}_{\alpha} [T_A](\vec \phi ')\,.\label{eq52}\end{equation}
The proof of Eq.\ (\ref{eq52}) is given in Appendix \ref{appenAbis}. 

Equation (\ref{eq52}) can also be obtained in the framework of fractional Fourier optics \cite{PPF4}. It has been deduced here from the previous analysis, based on ray transfers and homogeneous matrices.
Moreover, Eq.\ (\ref{eq52}) is similar to Eq.\ (\ref{eq46}), so that the spherical angular-spectrum propagation is accomplished by a fractional Fourier transformation of order $\alpha$, as well as the field-amplitude propagation \cite{SAS1,SAS2}.

%******************************************************
\section{Accordance with the Huygens--Fresnel principle}\label{sect6}
%******************************************************

\subsection{Expression with inhomogeneous variables}

\begin{figure}[b]%$$$$$$$$$$$$$$$$$$$$$$$$$$$$$$$$$$$$$$$$$$$$$$$$$$$$$$$$$$$$$$$$$$$
\begin{center}
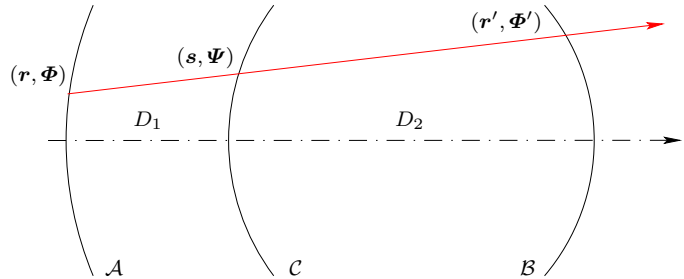
\end{center}
\caption{\small According to the Huygens principle, the transfer from ${\cal A}$ to ${\cal B}$ can split into the transfers from ${\cal A}$ to ${\cal C}$ and from ${\cal C}$ to ${\cal B}$.\label{fig8}}
\end{figure}%$$$$$$$$$$$$$$$$$$$$$$$$$$$$$$$$$$$$$$$$$$$$$$$$$$$$$$$$$$$$$$$$$$$$$$$$$$

In the framework of a scalar diffraction theory, the Huygens-Fresnel principle
states that the (electric) field-amplitude transfer from an emitter ${\cal A}$
to a receiver ${\cal B}$ can be split into the transfer from ${\cal A}$ to ${\cal C}$, followed by the transfer from ${\cal C}$ to ${\cal B}$, where ${\cal C}$ is an arbitrary  surface located between ${\cal A}$ and ${\cal B}$ \cite{Fog1}. We show in this section that the previous analysis, based on light rays and ray-matrices, is in accordance with the principle.

We use the previous notations and variables: $\vec r$ on ${\cal A}$ and $\vec r'$ on  ${\cal B}$, and we introduce an intermediate spherical cap ${\cal C}$ whose radius is $R_C$ (Fig.\ \ref{fig8}). A light ray on ${\cal C}$ is defined by $(\vec s,\vec \Psi)$. Let $D_1$ be the distance from ${\cal A}$ to ${\cal C}$ and $D_2$ the distance from ${\cal C}$ to ${\cal B}$, so that the distance from ${\cal A}$ to ${\cal B}$ is $D=D_1+D_2$. For the sake of convenience we assume ${\cal C}$ to be located between ${\cal A}$ and ${\cal B}$, but the analysis holds true for every ${\cal C}$; ${\cal C}$  might be a virtual emitter or receiver.

The spherical segment ${\cal C}$ is thought of as a receiver in the transfer from
${\cal A}$ to ${\cal C}$,  and as an emitter in the transfer from ${\cal C}$ to
${\cal B}$. It should be clear that the same pair $(\vec s,\vec \Psi)$, taken on ${\cal C}$, can be used
for describing both transfers.

 The transfer from ${\cal A}$ to ${\cal C}$ is described by
  \begin{equation}
   \begin{pmatrix}{\vec s\cr
     \vec \Psi }
   \end{pmatrix}
 =\begin{pmatrix}{
      \displaystyle{R_A-D_1\over R_A} & D_1 \cr \cr
      \displaystyle{D_1-R_A+R_C\over -R_AR_C} & \displaystyle{D_1+R_C\over R_C} }
    \end{pmatrix}
   \begin{pmatrix}{
     \vec r\cr 
     \vec \Phi }
   \end{pmatrix}
 \,,\label{eq72}\end{equation}
and the transfer from ${\cal C}$ to ${\cal B}$ by
\begin{equation}
   \begin{pmatrix}{
     \vec r'\cr 
     \vec \Phi'}
   \end{pmatrix}
 = \begin{pmatrix}{
      \displaystyle{R_C-D_2\over R_C} & D_2 \cr \cr
      \displaystyle{D_2-R_C+R_B\over -R_BR_C} & \displaystyle{D_2+R_B\over R_B}}
    \end{pmatrix}
   \begin{pmatrix}{
     \vec s\cr
     \vec \Psi }
   \end{pmatrix}
 \,.\label{eq73}\end{equation}
The combination of Eqs. (\ref{eq72}) and (\ref{eq73}) directly results in
  \begin{eqnarray}
   \begin{pmatrix}{
     \vec r'\cr
     \vec \Phi'}
   \end{pmatrix}
 &=&\begin{pmatrix}{
      \displaystyle{R_C-D_2\over R_C} & D_2 \cr \cr
      \displaystyle{D_2-R_C+R_B\over -R_BR_C} & \displaystyle{D_2+R_B\over R_B} }
    \end{pmatrix}
    \begin{pmatrix}{
      \displaystyle{R_A-D_1\over R_A} & D_1 \cr \cr
      \displaystyle{D_1-R_A+R_C\over -R_AR_C} & \displaystyle{D_1+R_C\over R_C} }
    \end{pmatrix}
   \begin{pmatrix}{
     \vec r\cr 
     \vec \Phi  }
   \end{pmatrix}
   \nonumber \\
   & &\nonumber \\
&=& \begin{pmatrix}{
      \displaystyle{R_A-D_1-D_2\over R_A} & D_1+D_2 \cr \cr
      \displaystyle{D_1+D_2-R_A+R_B\over -R_AR_B} & \displaystyle{D_1+D_2+R_B\over R_B} }
    \end{pmatrix}
   \begin{pmatrix}{
     \vec r\cr
     \vec \Phi}
   \end{pmatrix}
 \,,\label{eq75}\end{eqnarray}
and since $D=D_1+D_2$, Eq.\ (\ref{eq75}) is Eq. (\ref{eq26}) once more.

%*************************************************
\subsection{Expression with homogeneous variables}
%*************************************************
 
We examine now the accordance of the previous homogeneous ray-matrix representation with the Huygens-Fresnel principle. We have to compose two rotations whose angles are $-\alpha_1$ and $-\alpha_2$ and the result should be a rotation whose angle is $-\alpha =-\alpha_1-\alpha_2$. Nevertheless, the composition of the associated matrices physically makes sense  only if reduced variables on ${\cal C}$ are the same for both transfers. The problem has already been analyzed in  previous articles \cite{PPF4,Fog1} and the result is as follows: the composition makes sense if, and only if, the radius of ${\cal C}$ is $R_C$ such that
\begin{equation}
 R_C={D_1(D_2+R_B)(R_A-D)+D_2(D+R_B)(R_A-D_1)\over D_1(R_A-D)+D_2(D+R_B)}
 \,.\label{eq92}\end{equation}

Given an emitter ${\cal A}$ and a receiver ${\cal B}$, the result holds under rather strong conditions: (a) the field transfer from ${\cal A}$ to ${\cal B}$ is of real-order; (b) intermediate  caps (such as ${\cal C}$ above) belong to a family of  spherical caps, whose curvature radii take only specific values, according to Eq.\ (\ref{eq92}). Such a result is actually close to the historical way in which Huygens conceived light propagation.
Every point of an emitter emits `wavelets' in the form of spherical lightwaves, and the disturbance at a later instant is found on a wavefront, which is the envelope of the  wavelets. Each point of the wavefront, in turn, re-emits wavelets whose envelope at a later instant provides the wavefront where the disturbance can be found.
In this description of light propagation, an intermediate surface between an emitter and a receiver may not be an arbitrary cap, for the field on it has to correspond to an actual wavefront. Given a distance from the emitter, only one cap is then admissible, and in the metaxial theory, this cap is approximated as a sphere; its radius is given by Eq.\ (\ref{eq92}).

Such a situation corresponds to Gaussian beams \cite{PPF3}. A Gaussian beam  can be seen as a sequence of spherical wavefronts ${\cal W}_\alpha$ on which the electric-field amplitude is represented by a Gaussian function (or more generally a Hermite-Gauss function).  The field transfer from ${\cal W}_{\alpha_1}$ to ${\cal W}_{\alpha _2}$ ($\alpha_1<\alpha_2$) can be seen as the composition of two field transfers from ${\cal M}_{\alpha_1}$ to ${\cal M}_{\alpha_3}$  and from ${\cal M}_{\alpha_3}$ to ${\cal M}_{\alpha_2}$, where ${\cal M}_{\alpha_3}$ is an intermediate wavefront, belonging to the previous family of wavefronts that constitute the Gaussian beam.

%******************************************************
\section{Coherent imaging}\label{sect7}
%******************************************************

Let ${\cal S}$ be a centered system that forms the image ${\cal A}'$ of an arbitrary  spherical cap ${\cal A}$. A ray issued from a point $M$ on ${\cal A}$ is transformed into a ray on ${\cal A}'$ and the issue is to establish the relationship between the two rays. We consider geometrical images in the meaning that we do not take into account diffraction by limited apertures of lenses (or refracting spherical caps, or mirrors).

We first examine image formation by a refracting sphere before we generalize to an arbitrary centered system.

\subsection{Imaging by a refracting spherical cap}

We consider a refracting spherical cap ${\cal D}$ (radius $R_D$) separating two homogeneous and isotropic media with refractive indices $n$ and $n'$. We proceed in several steps that are as follows.

\subsubsection{Matrix form of Snell's law (refraction) \cite{PPF7}}

\begin{figure}[h]%$$$$$$$$$$$$$$$$$$$$$$$$$$$$$$$$$$$$$$$$$$$$$$$$$$$$$$$$$$$$$$$$$$$
\begin{center}
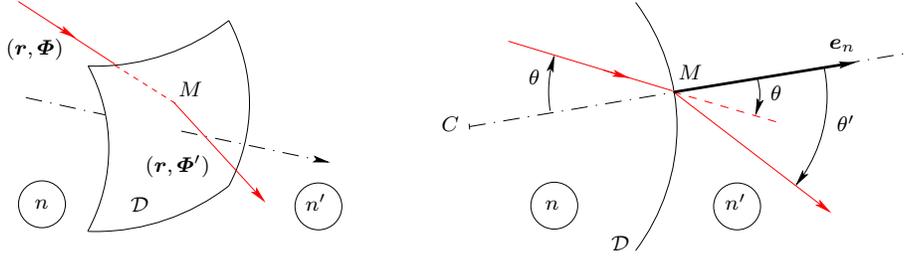
\end{center}
\caption{\small Snell's refraction-law. The incident ray $(\vec r,\vec \Phi )$ is refracted as ray $(\vec r,\vec \Phi ')$. Right: diagram in the plane of
incidence; C is the curvature center of the spherical cap ${\cal D}$; incident and refracted angles are taken from the normal at M towards the rays.\label{fig9}}
\end{figure}

A light ray $(\vec r,\vec  \Phi$) is incident on ${\cal D}$ at point $M$ (coordinates $\vec r$), see Fig.\ \ref{fig9}. The refracted ray is written $(\vec r ',\vec \Phi ')$ and since it passes through $M$, we have $\vec r'=\vec r$. If $\vec e_n$ denotes the unit vector normal to ${\cal D}$ at $M$, the incident angle $\theta$ is the angle taken from $\vec e_n$ to $\vec e_u$, and the refracted angle $\theta'$ is the angle from $\vec e_n$ to $\vec e'_u$, where $\vec e_u$ is along the incident ray and $\vec e'_u$ along the refracted ray (see Sect.\ \ref{sect2}). Since $||\vec \Phi ||=|\sin\theta|$, from  the second part of Snell's law ($n\sin\theta =n'\sin\theta '$), we obtain
\begin{equation}
  n||\vec \Phi ||=n'||\vec \Phi '||\,.\end{equation}
We assume $\vec \Phi \ne 0$. According to the first part of Snell's law, the incident and the refracted rays are in the plane of incidence, so that $\vec e_u$, $\vec e_u'$ and $\vec e_n$ are coplanar (Fig.\ \ref{fig9}, right). Then  $\vec \Phi$ and $\vec \Phi '$, which are along the respective projections of $\vec e_u$ and $\vec e_u'$ on  the plane tangent to ${\cal D}$ at $M$, are colinear, and since $\theta$ and $\theta '$ have the same sign, we obtain
\begin{equation}
  n\vec \Phi =n'\vec \Phi '\,,\end{equation}
which constitutes a vectorial form of  Snell's law, and which holds also for $\vec \Phi =0$.

Finally the matrix form of Snell's law is
\begin{equation}
  \begin{pmatrix}{\vec r '\cr\vec \Phi '}\end{pmatrix}=
  \begin{pmatrix}{1 & 0\cr \cr 0 & \displaystyle{n\over n'}}\end{pmatrix}
   \begin{pmatrix}{\vec r \cr\vec \Phi }\,,\label{eq68}\end{pmatrix}
\end{equation}
and holds for meridional as well as for skew rays (as encountered when the refracting surface has a rotational symmetry).

%$$$$$$$$$$$$$$$$$$$$$$$$$$$$$$$$$$$$$$$$$$$$$$$$$$$$$$$$$$$$$$$$$$$$$$$$$

\subsubsection{Ray transfer by a refracting spherical cap}\label{sect712}

Let ${\cal A}$ be an emitter at a distance $d$ from ${\cal D}$, in the object space;  and let ${\cal B}$ be  a receiver at a distance $d'$ from ${\cal D}$, in the image space (we choose notations that are currently used in geometrical optics: $d=\overline{O\Omega_A}$ and $d'=\overline{O\Omega_B}$, see Fig.\ \ref{fig10}; the diffraction distance to be taken into account for the transfer from ${\cal A}$ to ${\cal D}$ is $D=-d$). A ray $(\vec r,\vec \Phi )$ on ${\cal A}$ becomes $(\vec s,\vec \Psi )$ on ${\cal D}$, then $(\vec s',\vec \Psi ')=(\vec s,\vec \Psi ')$ after refraction, and eventually $(\vec r',\vec \Phi ')$ on ${\cal B}$. We use Eqs.\ ({\ref{eq26}) and  (\ref{eq68}) to obtain
\begin{equation}
   \begin{pmatrix}{
     \vec r'\cr
     \vec \Phi'}
   \end{pmatrix}
 =
    \begin{pmatrix}{
        \displaystyle{R_D-d'\over R_D} & d' \cr
        \cr
      \displaystyle{d'-R_D+R_B\over -R_BR_D} & \displaystyle{d'+R_B\over R_B} }
    \end{pmatrix}\!
\begin{pmatrix}{1 & 0\cr \cr 0 & \displaystyle{n\over n'}}\end{pmatrix}\!
    \begin{pmatrix}{
      \displaystyle{R_A+d\over R_A} & -d \cr \cr
      \displaystyle{d+R_A-R_D\over R_AR_D} & \displaystyle{R_D-d\over R_D} }
    \end{pmatrix}
   \begin{pmatrix}{
     \vec r\cr
     \vec \Phi}
   \end{pmatrix}\,,\label{eq54}
\end{equation}
from which we shall deduce the following results.

\begin{figure}%[h]%$$$$$$$$$$$$$$$$$$$$$$$$$$$$$$$$$$$$$$$$$$$$$$$$$$$$$$$$$$$$$$$
\begin{center}
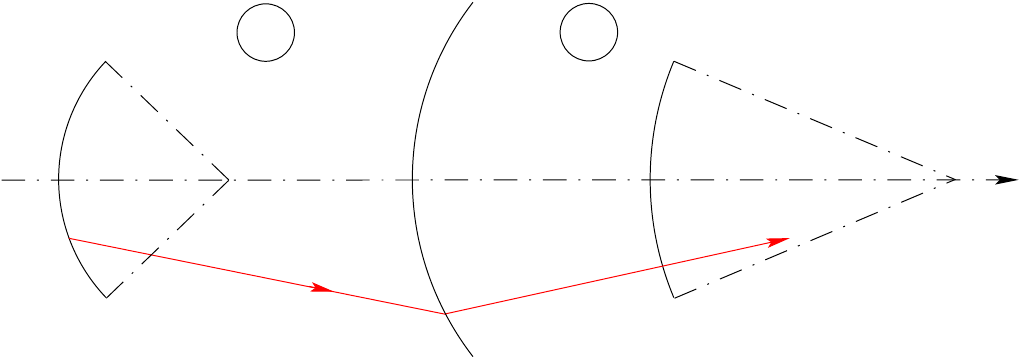
\end{center}
\caption{\small General transfer fron ${\cal A}$ to ${\cal B}$ with refraction on ${\cal D}$. \label{fig10}}
\end{figure}%$$$$$$$$$$$$$$$$$$$$$$$$$$$$$$$$$$$$$$$$$$$$$$$$$$$$$$$$$$$$$$$$$$$$

\subsubsection{Conjugation formula and lateral magnification}\label{sect73}%**********************
The receiver ${\cal B}$ becomes  the image ${\cal A}'$ of ${\cal A}$ if every ray passing through $\vec r$ passes through $\vec r'$ after having crossed the refracting surface, that is, if $\vec r'$ does not depend on $\vec \Phi$. According to eq.\ (\ref{eq54}), this happens if
\begin{equation}
-d{R_D-d'\over R_D}+{n\over n'}d'{R_D-d\over R_D}=0\,,\end{equation}
 that is
\begin{equation}
{n'\over d'}={n\over d}+{n'-n\over R_D}\,.\label{eq56}\end{equation}
Equation (\ref{eq56}) is a conjugation law of the refracting spherical cap.

From Eqs.\ (\ref{eq54}) and (\ref{eq56}) we deduce, in case of imaging,
\begin{equation}
\vec r'= \left[1 + {d'\over R_D}\left({n\over n'}-1\right)+{1\over R_A}\left(d-{n\over n'}d'-{dd'\over R_D}+{n\over n'}{dd'\over R_D}\right)\right]\vec r
= {nd'\over n'd}\,\vec r\,.\end{equation}
The imaging lateral-magnification is $m_{\rm v}=nd'/n'd$, a classical result of geometrical optics. (Subscript ``v'' indicates that $m_{\rm v}$ is the magnification between vertices of ${\cal A}$ and ${\cal A}'$.)

We deduce that Eq.\ (\ref{eq54}) can be written as
\begin{equation}
   \begin{pmatrix}{
     \vec r'\cr
     \vec \Phi'}
   \end{pmatrix}=
 \begin{pmatrix}{
     m_{\rm v} & 0\cr
     a_{21} & a_{22}}
   \end{pmatrix}
 \begin{pmatrix}{
     \vec r\cr
     \vec \Phi}
   \end{pmatrix}\,,\label{eq58}
\end{equation}
 where $a_{21}$ and $a_{22}$ remain to be determined.

\subsubsection{Determination of $a_{22}$}%*****************************************

From  Eqs.\ (\ref{eq54}) and (\ref{eq56}) we deduce
\begin{eqnarray}
a_{22}\!\!\!&=&\!\!\! d\,{d'-R_D+R_B\over R_DR_B}+{n\over n'}{(d'+R_B)(R_D-d)\over R_DR_B}\nonumber \\
&=&\!\!\!{1\over R_DR_B}\left[dd'-dR_D-{n\over n'}dd'+{n\over n'}d'R_D+R_B\left(d+{n\over n'}R_D-{n\over n'}d\right)\right]\nonumber \\
&=&\!\!\!{d\over d'}={n\over n'}{1\over m_{\rm v}}\,.
\end{eqnarray}

\subsubsection{Conjugation of curvature centers (double-conjugation law \cite{GB1,GB2})}%*********************************

The spherical receiver ${\cal A}'$ is the coherent geometrical image of the spherical
emitter ${\cal A}$ if the field amplitude on ${\cal A}'$ is equal to the field
amplitude on ${\cal A}$ to within a scaling factor which is equal to the
lateral magnification factor.  As a consequence, the phase is preserved in the imaging process: if $M$ and $N$ are two points on the spherical cap ${\cal A}$, the images of which are $M'$ and $N'$ on ${\cal A}'$, the phase difference between vibrations at $M'$ and $N'$ is equal to the phase difference between vibrations at $M$ and $N$.  The field amplitude on ${\cal A}'$ is related to the field amplitude on ${\cal A}$ by
\begin{equation}
  U_{A'}(\vec r')={1\over m_{\rm v}}\,U_A\left({\vec r'\over m_{\rm v}}\right)\,,\label{eq74}\end{equation}
where $m_{\rm v}$ is the lateral magnificationn at vertices: if $\Omega$ is the vertex of ${\cal A}$ (with $d=\overline{O\Omega}$) and $\Omega '$ the vertex of ${\cal A}'$  (with $d'=\overline{O\Omega '}$), points $\Omega$ and $\Omega '$ are conjugates and $d$ and $d'$ are linked by Eq.\ (\ref{eq56}). The factor $1/m_{\rm v}$ before $U_A$ is necessary to express that the power of the whole object is also the power of the whole image: $\int_{{\mathbb R}^2}|U_{A'}(\vec r')|^2\,\D \vec r'=\int_{{\mathbb R}^2}|U_{A}(\vec r)|^2\,\D \vec r$.

Let ${\cal F}$ be the Fourier sphere of ${\cal A}$, so that  
\begin{equation}
  U_F(\vec s)={\I\over \lambda R_A}\,\widehat U_A\left({\vec s\over \lambda R_A}\right)\,.\end{equation}
From Eq.\ (\ref{eq74}) we deduce
\begin{equation}
  \widehat U_{A'}(\vec F)=m_{\rm v}\,\widehat U_A(m_{\rm v}\vec F)\,,\end{equation}
and if ${\cal F}'$ denotes  the Fourier sphere of ${\cal A}'$ we have
\begin{equation}
  U_{F'}(\vec s')={\I\over \lambda ' R_{A'}}\,\widehat U_{A'}\left({\vec s'\over \lambda 'R_{A'}}\right)
  ={\I m_{\rm v}\over \lambda 'R_{A'}}\;\widehat U_{A}\left({m_{\rm v}\vec s'\over \lambda 'R_{A'}}\right)
  =m_{\rm v}{\lambda R_A\over \lambda 'R_{A'}}\;U_F\left(m_{\rm v}{\lambda R_A\vec s'\over \lambda 'R_{A'}}\right)\,.\label{eq77n}\end{equation}
Equation (\ref{eq77n}) has the form $U_{F'}(\vec s')= (1/m)U_F(\vec s'/m)$, that is, the form of Eq.\ (\ref{eq74}): we conclude that the field amplitude on ${\cal F}'$ is the coherent image of the field amplitude on ${\cal F}$, which means that the spherical cap ${\cal F}'$ is the coherent image of ${\cal F}$. The vertices of ${\cal F}$ and ${\cal F}'$, say $C$ and $C'$, are conjugates. Since $C$ is also the curvature center of ${\cal A}$ and $C'$ the curvature center of ${\cal A}'$, we conclude that the spherical cap ${\cal A}'$ is the coherent image of the spherical cap ${\cal A}$ if, and only if,
\begin{itemize}
\item the vertex of ${\cal A}'$ is the paraxial image of the vertex of ${\cal A}$;
\item the curvature center of ${\cal A}'$ is the paraxial image of the curvature center of ${\cal A}$.
\end{itemize}
That constitutes  the ``double-conjugation law'' of geometrical coherent-imaging for a refracting spherical-cap (a refracting plane constitutes a particular case: the curvature radius is infinite).  The law also holds for mirrors and can be generalized to every centered system made up of refracting spheres and mirrors \cite{PPF3,PPF5,GB1,GB2}.

In Appendix B, we provide a pure geometrical proof of the conjugation of curvature centers for imaging by a refracting spherical-cap (another geometrical proof has been given in a recent article \cite{PPF5}).

\subsubsection{Determination of $a_{21}$}

We consider a ray $(\vec r,\vec \Phi )$ on ${\cal A}$, such that $\vec \Phi =0$. Then the ray passes through $C$ (the curvature center of ${\cal A}$) and its image ray $(\vec r' ,\vec \Phi ')$ should pass through  $C'$, the center of ${\cal A}'$, so that $\vec \Phi '=0$. Since the results holds true for every $\vec r$, we must have $a_{21}=0$. (We provide an analytic checking of this result in Appendix C.)

Finally we arrive at
\begin{equation}
   \begin{pmatrix}{
     \vec r'\cr
     \vec \Phi'}
   \end{pmatrix}=
 \begin{pmatrix}{
     m_{\rm v} & 0\cr \cr
     0 & \displaystyle{1\over m_{\rm v}}{n\over n'}}
   \end{pmatrix}
 \begin{pmatrix}{
     \vec r\cr
     \vec \Phi}
   \end{pmatrix}\,.\label{eq79n}
\end{equation}

We remark that
\begin{equation}
n\,\vec r\vec \cdot\vec \Phi =n'\,\vec r'\vec \cdot\vec \Phi'\,,\end{equation}
and since $n\lambda =n'\lambda '$, from $\vec \Phi =\lambda \vec F$ we deduce
\begin{equation}
\vec r\vec \cdot\vec F=\vec r'\vec \cdot\vec F'\,.\end{equation}

\subsubsection{Radius magnification}\label{sect77}%**********************************
Let ${\cal A}'$ (vertex $\Omega '$, center $C'$, radius $R_{A'}=\overline{\Omega 'C'}$) be the coherent image of ${\cal A}$ (vertex $\Omega$, center $C$, radius $R_A=\overline{\Omega C}$) through a refracting spherical cap ${\cal D}$ (vertex $O$, radius $R_D$). Let us denote $q=\overline{OC}=d+R_A$ and $q=\overline{OC'}=d'+R_{A'}$. Since $C$ and $C'$ are conjugates, we have
\begin{equation}
  {n'\over q'}={n\over q}+{n'-n\over R_D}\,,\end{equation}
and the corresponding lateral magnification is
\begin{equation}
  m_{\rm c}={nq'\over n'q}\,.\end{equation}
From Eq.\ (\ref{eq77n}) we also deduce
\begin{equation}
  m_{\rm c}={\lambda 'R_{A'}\over m_{\rm v}\lambda R_A}={1\over m_{\rm v}}\,{n\over n'}\,{R_{A'}\over R_A} %={nd'^2R_{A'}\over n'd^2R_A}
  \,,
\end{equation}
and then
\begin{equation}
  m_{\rm r}={R_{A'}\over R_A}={n'\over n}\,m_{\rm v}\,m_{\rm c}\,,\label{eq91}\end{equation}
where $m_{\rm r}$ is called the radius magnification \cite{PPF3,GB1,GB2}.

The radius magnification-law can also be deduced from vertex and center conjugation-formulas, as shown in Appendix C.

\subsection{Generalization to centered systems}%***************************************************

A centered system ${\cal S}$ is the succession of refracting spherical caps ${\cal D}_i$, $i=1,...,I$, where ${\cal D}_i$ separates two media with respectives indices $n_{i-1}$ and $n_i$. We denote $n_0=n$ and $n_I=n'$. Let ${\cal A}_0$ be an object (optically located in the medium of index $n$) and ${\cal A}_i$ (vertex $\Omega_i$, center $C_i$) be the intermediate image, that is, the image of ${\cal A}_{i-1}$ through ${\cal D}_i$. We denote ${\cal A}'={\cal A}_I$, the final image (optically located in the medium of index $n'$). We apply the double conjugation law:
\begin{itemize}
\item Since $\Omega_i$ is the paraxial image of $\Omega_{i-1}$ through ${\cal D_i}$,  we obtain that $\Omega '=\Omega_I$ is the paraxial image of $\Omega_0=\Omega$ through ${\cal S}$;
\item Since $C_i$ is the paraxial image of $C_{i-1}$ through ${\cal D_i}$,  we obtain that $C'=C_I$ is the paraxial image of $C_0=C$ through ${\cal S}$.
\end{itemize}
We conclude that the double-conjugation law applies to ${\cal S}$.

A ray $(\vec r,\vec \Phi )=(\vec r_0,\vec \Phi_0)$, issued from ${\cal A}_0$, becomes $(\vec r_i,\vec \Phi_i)$ on ${\cal A}_i$, after refraction on ${\cal D}_i$. The final ray, on ${\cal A}'$,  is $(\vec r',\vec\Phi ')=(\vec r_I,\vec \Phi_I)$.

For every $i$ we have $\vec r_i=m_i \vec r_{i-1}$, so that
\begin{equation}
  \vec r'=\vec r_I=m_{I}\vec r_{I-1}=m_{I}m_{I-1} \vec r_{I-2}=\cdots = m_{I}m_{I-1}\cdots m_{1}\vec r_{0}=m_{\rm v}\vec r\,,
\end{equation}
where $m_{\rm v}=\displaystyle{\prod_{i=1}^{i=I} m_i}$ is the lateral magnification for the conjugation of $\Omega$ and $\Omega '$.
We also have
\begin{equation}
  {n\over n'}={n\over n_1}\;{n_1\over n_2}\cdots {n_{I-2}\over n_{I-1}}\;{n_{I-1}\over n'}\,.\end{equation}
Equation (\ref{eq79n}) leads us to write
\begin{equation}
   \begin{pmatrix}{
     \vec r'\cr
     \vec \Phi'}
   \end{pmatrix}=\begin{pmatrix}{
     \vec r_I\cr
     \vec \Phi_I}
   \end{pmatrix}=\prod_{i=1}^{i=I}
 \begin{pmatrix}{
     m_i & 0\cr \cr
     0 & \displaystyle{1\over m_{i}}{n_{i-1}\over n_i}}
   \end{pmatrix}
 \begin{pmatrix}{
     \vec r_0\cr
     \vec \Phi_0}
 \end{pmatrix}
 =
  \begin{pmatrix}{
     m_{\rm v} & 0\cr \cr
     0 & \displaystyle{1\over m_{\rm v}}{n\over n'}}
   \end{pmatrix}
 \begin{pmatrix}{
     \vec r\cr
     \vec \Phi}
 \end{pmatrix}\,,
 \label{eq88}
\end{equation}
because all previous square matrices are diagonal. Equation (\ref{eq88}) is the generalization of Eq.\ (\ref{eq79n}) to a centered system made up of refracting spherical caps. It can be proved to hold also for catadioptric systems.

The radius magnification law is
\begin{equation}
  m_{\rm r}={R_{A'}\over R_A}={R_I\over R_0}={R_I\over R_{I-1}}\,{R_{I-1}\over R_{I-2}}\cdots {R_2\over R_1}\,{R_1\over R_0}=\prod_{i=1}^{i=I}{n_i\over n_{i-1}}\, m_{{\rm v},i}\,m_{{\rm c},i}={n'\over n}\,m_{\rm v}\,m_{\rm c}\,,
\end{equation}
and  takes the same form as Eq.\ (\ref{eq91}).

\subsection{Homogeneous matrix representation}

Since ${\cal F}_0$ is the identity operator (${\cal F}_0[f]=f$, for every function $f$) and since ${\cal F}_\pi [f](\vec
\rho ')= f(-\vec \rho ')$, the imaging should be expressed by a fractional Fourier
transform whose order is $0$ or iqual to $\pm\pi$.  That holds true for both reduced spatial variable $\vec \rho$ and angular frequency $\vec\phi$, because the field transfer and the angular-spectrum transfer are both expressed by fractional Fourier transforms of equal orders.  The homogeneous  ray-matrix associated with imaging thus results to be such that
\begin{equation}
  \begin{pmatrix}{\vec \rho ' \cr \vec\phi '}\end{pmatrix} =
\pm \begin{pmatrix}{
1 & 0 \cr  0 & 1} \begin{pmatrix}{\vec \rho \cr \vec\phi }\end{pmatrix}
\end{pmatrix}\,.\label{eq96}
\end{equation}
In the appendix E, we prove that this is the case.

\medskip \noindent {\bf  Remark 1.} Square matrices in Eq.\ (\ref{eq88}) are homogenous since $m_{\rm v}$ and refractive indices $n$ and $n'$ are pure numbers. But column-vectors are not homogeneous. In Eq.\ (\ref{eq96}) all matrices are homogeneous.

\medskip \noindent {\bf  Remark 2 (Reduced form of Snell's law).} Snell's law of refraction is expressed by Eq.\ (\ref{eq68}), which may be seen like an imaging between ray $(\vec r,\vec \Phi )$ and   ray $(\vec r',\vec \Phi ')$, with $\vec r'=\vec r$, so that the lateral magnification is $m_{\rm v}=1$. Then we have $\vec \phi '=\vec \phi$, which constitutes the reduced form of Snell's law.

%*****************************************************
\section{Conclusion}\label{conc}
%*****************************************************

Fractional Fourier optics is based on the representation of a Fresnel diffraction phenomenon by a fractional-order Fourier transformation. The transformation associated with a given diffraction phenomenon has been deduced here from a matrix representation of ray transfer from a spherical emitter to a spherical receiver by looking for homogeneous transfer matrices.  When the field transfer is expressed by a real-order transformation, the ray matrix is a 4--dimensional rotation matrix that splits into two rotations operating on 2-dimensional subspaces of the reduced phase space. The analysis can be extended to complex orders, the previous rotation matrices become then  2--dimensional hyperbolic-rotation  matrices \cite{PPF6}.

\section*{Appendix A. Proof of Eq.\ (\ref{eq45}) }\label{appenA}%**************************************
Changing variables in Eq.\ (\ref{eq42}) are as follows.

\smallskip
\noindent(i) We begin with
\begin{equation}
  {\vec r\vec\cdot \vec r'\over \lambda D}={\vec\rho\vec\cdot\vec \rho '\over \lambda DAA'}\,,
\end{equation}
with
\begin{equation}
  (AA')^4={1\over \lambda^4{R_A}^2{R_B}^2}\;{(D-R_A+R_B)^2\over D^2}\,.\end{equation}
We have
\begin{equation}
  {1\over\sin^2\alpha} =1+\cot^2\alpha =1+{(R_A-D)(R_B+D)\over D(D-R_A+R_B)}={R_AR_B\over D(D-R_A+R_B)}\,,\label{eq97}
\end{equation}
and since $\alpha$ has the sign of $D$, we obtain
\begin{equation}
  {\vec r\vec\cdot \vec r'\over \lambda D}={\vec \rho\vec\cdot\vec\rho '\over \sin\alpha}\,.\end{equation}

\smallskip
\noindent(ii) Then we consider
\begin{equation}
  {1\over\lambda}\left({1\over D}-{1\over R_A}\right)r^2={1\over \lambda}\;{R_A-D\over DR_A}\,{\rho^2\over A^2}\,.\end{equation}
We have
\begin{equation}
{1\over\lambda^2}\;{(R_A-D)^2\over D^2{R_A}^2}{1\over A^4}={(R_A-D)(R_B+D)\over D(D-R_A+R_B)}=\cot^2\alpha\,,
\end{equation}
and eventually, since the sign of $\cot\alpha$ is the sign of $R_AD(R_A-D)$,
\begin{equation}
  {1\over\lambda}\left({1\over D}-{1\over R_A}\right)r^2=\rho^2\cot\alpha\,.\end{equation}

\smallskip
\noindent(iii) We have
\begin{equation}
  {1\over\lambda}\left({1\over D}+{1\over R_A}\right)r'^2={1\over \lambda}\;{R_B+D\over DR_B}\,{\rho'^2\over A'^2}\,.\end{equation}
and then
\begin{equation}
{1\over\lambda^2}\;{(R_B+D)^2\over D^2{R_B}^2}{1\over A'^4}={(R_A-D)(R_B+D)\over D(D-R_A+R_B)}=\cot^2\alpha \,.
\end{equation}
Since the sign of $\cot\alpha$ is the sign of  $R_BD(D+R_B)$, we obtain
\begin{equation}
  {1\over\lambda}\left({1\over D}+{1\over R_B}\right)r'^2=\rho'^2\cot\alpha\,.\end{equation}

\smallskip
\noindent(iv) Since both $\vec r$ and $\vec \rho$ are two-dimensional vectors, we have
\begin{equation}
  {\D\vec r\over \lambda D}={1\over \lambda D}\;{\D\vec\rho\over A^2}\,.\end{equation}
We have
\begin{eqnarray}
  \lambda^2D^2A^2={D^2\over {R_A}^2}\;{(R_A-D)(D-R_A+R_B)\over D(R_B+D)}
  &=&{D(D-R_A+R_B)\over R_AR_B}\;{R_A-D\over R_B+D}\;{R_B\over R_A}\nonumber \\
  &=&{R_B(R_A-D)\over R_A(R_B+D)}\sin^2\alpha\,. 
\end{eqnarray}
We note that $(R_B/R_A)(R_A-D)/(R_B+D)\ge 0$, because $R_AD(R_A-D)$ and $R_BD(D+R_B)$ have the same sign.
Since $\alpha$ has the sign of $D$, we obtain
\begin{equation}
  {\D\vec r\over \lambda D}={1\over \sin\alpha}\,\sqrt{R_A(R_B+D)\over R_B(R_A-D)}\,\D\vec \rho\,.
\end{equation}

Making the changes of the above four items lead to write Eq.\ (\ref{eq42}) in the form of Eq.\ (\ref{eq45}).

\section*{Appendix B. Proof of Eq.\ (\ref{eq52}) }\label{appenAbis}%**************************************

We use reduced angular-frequencies and reduced angular-spectra to write Eq.\ (\ref{eq61s}) as a fractional Fourier transform of order $\alpha$. Reduced angular frequencies are $\vec \phi =B\vec \Phi$ and $\vec \phi '=B'\,\vec \Phi '$, where $B$ and $B'$ are positive and  given by Eqs.\ (\ref{eq37}) and (\ref{eq39}). Reduced angular-spectra are $T_A$ and $T_B$, given by Eqs.\ (\ref{eq63s}) and (\ref{eq64s}).

\smallskip
\noindent (i) We begin with the exponential depending on $\vec \Phi\vec \cdot \vec \Phi '$. We have
\begin{equation}
  {R_AR_B\over \lambda (D-R_A+R_B)}\; \vec \Phi\vec \cdot \vec \Phi '= {R_AR_B\over \lambda (D-R_A+R_B)}\;{\vec \phi\vec\cdot \vec\phi '\over BB'}\,,\end{equation}
with
\begin{equation}
  (BB')^4={R_A^2\,R_B^2\over \lambda^4}\;{D^2\over (D-R_A+R_B)^2}\,.
  \end{equation}
Since $R_AR_BD(D-R_A+R_B)$ is positive according to Eq.\ (\ref{eq97}), we have
\begin{equation}
  (BB')^2={R_A\,R_B\over \lambda^2}\;{D\over (D-R_A+R_B)}\,.
\end{equation}
Then,  according to Eq.\ (\ref{eq97}) once more, 
\begin{equation}
  \left[{R_AR_B\over \lambda (D-R_A+R_B)}\right]^2\;{1\over (BB')^2}={R_AR_B\over D(D-R_A+R_B)}={1\over \sin^2\alpha}\,.
\end{equation}
Since $R_AR_BD(D-R_A+R_B)$ is positive and since $\alpha$ (and $\sin\alpha$) has the sign of $D$, we conclude that $R_AR_B(D-R_A+R_B)$ and then $R_AR_B(D-R_A+R_B)BB'$  also have the sign of $\alpha$, so that eventually
\begin{equation}
 {R_AR_B\over \lambda (D-R_A+R_B)}\;\vec\Phi\vec\cdot\vec \Phi'={1\over \sin\alpha}\;\vec\phi\vec\cdot\vec\phi '\,.
\end{equation}

\smallskip
\noindent(ii) Factor in $\Phi^2$. We have
\begin{equation}
  {R_A(R_B+D)\over \lambda (D-R_A+R_B)}\,\Phi^2={R_A(R_B+D)\over \lambda (D-R_A+R_B)}\,{\phi^2\over B^2}\,,\end{equation}
and then
\begin{equation}
  \left[{R_A(R_B+D)\over \lambda (D-R_A+R_B)}\right]^2\,{1\over B^4}={(R_A-D)(D+R_B)\over D(D-R_A+R_B)}=\cot^2\alpha\,.
\end{equation}
Since $R_AR_B D (D-R_A+R_B)$ is positive (see above), we conclude that $R_A(D-R_A+R_B)$ and $R_BD$ have the same sign. On the other hand, $R_BD(R_B+D)$ has the sign of $\cot \alpha$ (as shown in Sect. \ref{sect31}). We conclude that  $R_A(R_B+ D) (D-R_A+R_B)$ and $\cot\alpha$ have the same sign, and we may write
\begin{equation}
  {R_A(R_B+D)\over \lambda (D-R_A+R_B)}\,{1\over B^2}=\cot\alpha\,,
\end{equation}
so that
\begin{equation}
  {R_A(R_B+D)\over \lambda (D-R_A+R_B)}\,\Phi^2=\phi^2\cot\alpha\,,\end{equation}

\smallskip
\noindent(iii) Factor in ${\Phi'}^{2}$. We have
\begin{equation}
  {R_B(R_A-D)\over \lambda (D-R_A+R_B)}\,{\Phi '}^2={R_B(R_A-D)\over \lambda (D-R_A+R_B)}\,{{\phi '}^2\over {B'}^2}\,,\end{equation}
and then
\begin{equation}
  \left[{R_B(R_A-D)\over \lambda (D-R_A+R_B)}\right]^2\,{1\over {B'}^4}={(R_A-D)(D+R_B)\over D(D-R_A+R_B)}=\cot^2\alpha\,.
\end{equation}
As above, we show that $R_B(R_A- D) (D-R_A+R_B)$ and $\cot\alpha$ have the same sign, and we eventually obtain
\begin{equation}
  {R_B(R_A-D)\over \lambda (D-R_A+R_B)}\,{\Phi '}^2={\phi '}^2\cot\alpha\,.\end{equation}

\smallskip
\noindent (iv) Differential term. Since both $\vec \Phi$ and $\vec \phi$ are 2--dimensional variables, we have
\begin{equation}
  {R_AR_B\over \lambda (D-R_A+R_B)}\,\D \vec\Phi = {R_AR_B\over \lambda (D-R_A+R_B)}\,{1\over B^2}\,\D\vec\phi\,.\end{equation}
Then
\begin{equation}
  \left[{R_AR_B\over \lambda (D-R_A+R_B)}\right]^2\,{1\over B^4}={R_B^2(R_A-D)\over D(R_B+D)(D-R_A+R_B)}
  ={R_B(R_A-D)\over R_A(R_B+D)}\;{1\over \sin^2\alpha}\,.\end{equation}
Since $R_AR_B(D-R_A+R_B)$ has the sign of $\alpha$ (see item (i) above), and since  $R_A(R_A-D)$ and $R_B(R_B+D)$ have the same sign, we conclude
\begin{equation}
  {R_AR_B\over \lambda (D-R_A+R_B)}\,\D \vec\Phi = \sqrt{R_B(R_A-D)\over R_A(R_B+D)}\;{\D\vec \phi\over \sin\alpha}\,.\end{equation}

\smallskip
\noindent (v) The previous changes of variables lead us to write Eq.\ (\ref{eq61s}) in the form
\begin{eqnarray}
  S_B\left({\vec \phi '\over B'}\right)\!\!\!\!&=&\!\!\!\! {\I\over \sin\alpha}\,\sqrt{R_B(R_A-D)\over R_A(R_B+D)}
  \exp (-\I\pi {\phi '}^2\cot\alpha) \nonumber \\
  & &\hskip 1cm\times
  \int_{{\mathbb R}^2}\exp (-\I\pi \phi^2\cot\alpha)\,\exp\left({2\I\pi \vec \phi\vec\cdot\vec\phi' \over \sin\alpha}\right)\;S_A\left({\vec\phi \over B}\right)\,\D\vec\phi\,,
\end{eqnarray}
that is
\begin{eqnarray}
  T_B(\vec\phi ')\!\!\!\!&=&\!\!\!\! \sqrt{\displaystyle\left|R_B+D\over R_BD\right|}\;
  S_B\left(\displaystyle{\vec \phi '\over B'}\right)\nonumber \\
  &=&\!\!\!\! {\I\over \sin\alpha}  \exp (-\I\pi {\phi '}^2\cot\alpha)
  \int_{{\mathbb R}^2}\exp (-\I\pi \phi^2\cot\alpha) \nonumber \\
  & &\hskip 2cm\times
  \exp\left({2\I\pi \vec \phi\vec\cdot\vec\phi \over \sin\alpha}\right)\;\sqrt{\left|R_A-D\over R_AD\right|}\;
  S_A\left({\vec\phi \over B}\right)\,\D\vec\phi \nonumber \\
  \!\!\!\!&=&\!\!\!\!
  {\I\over \sin\alpha}  \exp (-\I\pi {\phi '}^2\cot\alpha)
  \int_{{\mathbb R}^2}\exp (-\I\pi \phi^2\cot\alpha)
  \exp\left({2\I\pi \vec \phi\vec\cdot\vec\phi' \over \sin\alpha}\right)
  T_A(\vec\phi )\,\D\vec\phi  \nonumber \\
    \!\!\!\!&=&\!\!\!\! \E^{\I\alpha}{\cal F}_\alpha [T_A](\vec\phi ')\,.
\end{eqnarray}

\section*{Appendix C. An alternative proof of the  conjugation of curvature centers }\label{appenB}%******

Let ${\cal A}'$ (center $C'$) be the coherent image of ${\cal A}$ (center $C$)  through the refracting spherical cap ${\cal D}$ (Fig.\ \ref{fig11}). Let $M$ and $N$ be two points on ${\cal A}$ and let $M'$ and $N'$ be their images on ${\cal A}'$.

According to the Fermat's principle, the optical path from $M$ to $M'$ is a constant for every light ray passing through $M$ and $M'$ and we can speak of the optical path $[MM']$. (That is rigorous if $M$ and $M'$ are stigmatic points, and holds up to second order in case of approximate stigmatism.) The same holds for the optical path $[NN']$.

Since ${\cal A}'$ is the coherent image of ${\cal A}$, the phase difference between vibrations at $M'$ and $N'$ is equal to the phase difference between vibrations at $M$ and $N$. If $N$ tends to $M$, then $[NN']$ tends to $[MM']$ and by continuity we otain $[NN']=[MM']$ for every pair $(M,N)$, where $M$ and $N$ belong to ${\cal A}$.

We then consider the optical path $[MCM']$, which intersects ${\cal D}$ at $L$, and the optical path $[NCN']$, which intersects ${\cal D}$ at $K$. We have
\begin{equation}
  [MLM']=[MCLM']=[MM']=[NN']=[NCKN']=[NKN']\,.\label{eq104}\end{equation}
Since  $C$ is the center of curvature of ${\cal A}$ and $C'$ the center of curvature of  ${\cal A}'$, we have
$[CM]=[CN]$ and $[M'C']=[N'C']$ so that, by Eq.\ (\ref{eq104}), we obtain
\begin{equation}
  [CLC']=[CM]+[MLM']+[M'C']=[CN]+[NKN']+[N'C']=[CKC']\,.\end{equation}
When $M$ and $N$ describe ${\cal A}$, points $K$ and $L$ describe ${\cal D}$, and we have $[CLC']=[CKC']$, which means that whatever $L$, the optical path $[CLC']$ is constant, so that $C'$ is the image of $C$: curvature centers of ${\cal A}$ and ${\cal A}'$ are conjugates. The proof is complete.

 \begin{figure}[h]%$$$$$$$$$$$$$$$$$$$$$$$$$$$$$$$$$$$$$$$$$$$$$$$$$
\begin{center}
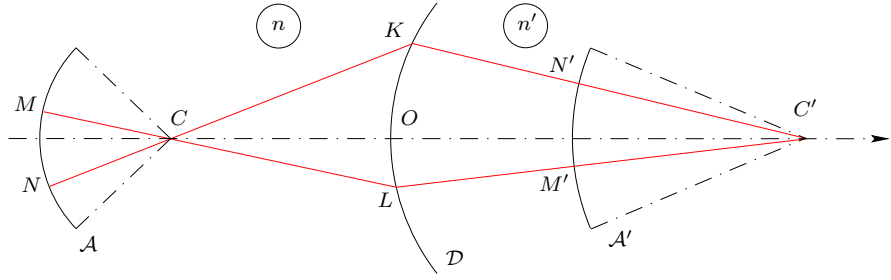
\end{center}
\caption{\small If ${\cal A}'$ (center $C'$) is the coherent image of ${\cal A}$ (center $C$), then $C'$ is necessarily the image of $C$.\label{fig11}}
\end{figure}%$$$$$$$$$$$$$$$$$$$$$$$$$$$$$$$$$$$$$$$$$$$$$$$$$$$$$$$

\section*{Appendix D. Checking $a_{21}=0$}\label{appenC}%***********

We refer to notations of Sect.\ \ref{sect712}. If $C$ is the center of ${\cal A}$, we denote $q=\overline{OC}=d+R_A$, and if $C'$ is the center of ${\cal B}$, we denote $q'=\overline{OC'}=d'+R_B$.

According to Eq.\ (\ref{eq54}) we have
\begin{equation}
  a_{21}=-{R_A+d\over R_A}\;{d'-R_D+R_B\over R_BR_D}+{n\over n'}\;{d'+R_B\over R_AR_BR_D}\;(d+R_A-R_D)={\frak{N}\over n'R_AR_BR_D}\,,
\end{equation}
where
\begin{equation}
  \frak{N}=-n'(R_A+d)(d'-R_D+R_B)+n(d'+R_B)(d+R_A-R_D)\,.
\end{equation}
We have 
\begin{equation}
  \frak{N}=-n'q(q'-R_D)+nq'(q-R_D)=qq'R_D\left({n'\over q'}-{n\over q}-{n'-n\over R_D}\right)\,.
\end{equation}
If $C'$ is the paraxial image of $C$, the conjugation formula gives $(n'/q')=(n/q)+[(n'-n)/R_D]$ and $\frak{N}=0$, so that eventually $a_{21}=0$.
  
\section*{Appendix E. An alternative proof of the radius magnification law}\label{appenD}%************
In this appendix, we directly deduce the radius magnification law (refracting sphere) from conjugation formulae for vertices and curvature centers.
We use notations of Sects.\ \ref{sect73} and \ref{sect77}. Since $\Omega $ and $\Omega '$ are conjugates, and since $C$ and $C'$ are also conjugates, we have
  \begin{equation}
{n'\over d'}-{n\over d}={n'-n\over R_D}={n'\over q'}-{n\over q}\,.\label{eq65}\end{equation}
The lateral magnification between $\Omega$ and $\Omega '$ is $m_{\rm v}=nd'/n'd$, and the
lateral  magnification between $C$ and $C'$ is $m_{\rm c}=nq'/n'q$.

The radius magnification is
\begin{equation}
m_{\rm r}={R_{A'}\over R_A}\,,\end{equation}
and we deduce from Eq.\ (\ref{eq65}) %and (\ref{eq66})
\begin{equation}
m_{\rm r}={q'-d'\over q-d}={nq'd'\over n'qd}={n'\over n}m_{\rm v}m_{\rm c}\,.\end{equation}

\section*{Appendix F. Homogeneous imaging-matrix}\label{appenE}%*******************************************************************
Let ${\cal D}$ (vertex $O$, radius $R_D$) be a refracting spherical cap, separating two media of refractive indices $n$ and $n'$ (corresponding wavelengths are $\lambda$ and $\lambda '$, and $n\lambda =n'\lambda '$). Let ${\cal A}$ (vertex $\Omega$, center $C$, radius $R_A=\overline{\Omega C}$) be a spherical emitter in the object space and let ${\cal A}'$ (vertex $\Omega '$, center $C'$, radius $R_{A'}=\overline{\Omega 'C'}$)
be its coherent image through the refracting surface ${\cal D}$.
We use notations of  Sects.\ \ref{sect73} and \ref{sect77}: $d=\overline{O\Omega}$, $d'=\overline{O\Omega '}$, $q=\overline{OC}=d+R_A$, $q'=\overline{OC'}=d'+R_{A'}$.
%\begin{equation}
%q=R_A+d\,,\hskip .5 cm q'=R_{A'}+d'\,,\end{equation}
%and because of the twofold conjugaion law, we have
%\begin{equation}
%{n'-n\over R_D}={n'\over d'}-{n\over d}={n'\over q'}-{n\over
%  d}\,.\end{equation}

\bigskip

\noindent {\it (i) Composition of transformations}

\medskip

From Eq.\ (\ref{eq65}) we deduce
\begin{equation}
d'={n'dR_D\over nR_D+d(n'-n)}\,,\hskip 1cm
q'={n'qR_D\over nR_D+q(n'-n)}\,,\label{eq110}\end{equation}
and then
\begin{equation}
R_D-d'={nR_D (R_D-d)\over nR_D+d(n'-n)}\,,\hskip 1cm
R_D-q'={nR_D (R_D-q)\over nR_D+q(n'-n)}\,.\label{eq111}\end{equation}

We choose coordinates $\vec r$ on ${\cal A}$, $\vec s$ on ${\cal D}$ and $\vec r'$ on ${\cal A}'$.
According to Eq.\ (\ref{eq38}), since the diffraction distance is $D=-d$, the transfer from
${\cal A}$ to ${\cal D}$ is expressed by choosing the following  reduced space
variable on ${\cal D}$
\begin{equation}
\vec \sigma =\left[{(R_D-d)(d+R_A-R_D)\over
    \lambda^2R_D^2 d(R_A+d)}\right]^{1/4}\vec s
=\left[{(R_D-d)(q-R_D)\over \lambda^2R_D^2dq}\right]^{1/4}\vec s\,,
\end{equation}
and according to Eq.\ (\ref{eq36}), the reduced variable on ${\cal D}$ corresponding to the transfer from ${\cal D}$ to ${\cal A}'$ (the diffraction distance is $D=d'$) is
\begin{equation}
\vec \sigma '=\left[{(R_D-d')(d'-R_D+R_{A'})\over
    \lambda'^2R_D^2 d'(d'+R_{A'})}\right]^{1/4}\vec s
=\left[{(R_D-d')(q'-R_D)\over \lambda'^2R_D^2d'q'}\right]^{1/4}\vec
    s\,.\end{equation}
By Eqs.\ (\ref{eq110}) and (\ref{eq111}) we conclude that $\vec \sigma =\vec
\sigma '$.

The angular frequencies on ${\cal D}$ are $\vec \Psi$ and $\vec \Psi '$, with $n\vec \Psi=n'\vec\Psi '$ (Snell's law). The corresponding reduced angular-frequencies are
\begin{equation}
\vec \psi =\left[{R_D^2d(R_A+d)\over
    \lambda^2 (R_D-d)(d+R_A-R_D)}\right]^{1/4}\vec \Psi
=\left[{R_D^2dq\over \lambda^2(R_D-d)(q-R_D)}\right]^{1/4}\vec \Psi \,,
\end{equation}
\begin{equation}
\vec \psi '=\left[{R_D^2d'(d'+R_{A'})\over
    \lambda'^2 (R_D-d')(d'-R_D+R_{A'})}\right]^{1/4}\vec \Psi '
=\left[{R_D^2d'q'\over \lambda'^2(R_D-d')(q'-R_{D})}\right]^{1/4}\vec \Psi ' \,,
\end{equation}
and since $n\vec \Psi =n'\vec \Psi '$, by Eqs.\ (\ref{eq110}) and (\ref{eq111}), we obtain: $\vec \psi =\vec \psi '$.

The ray transfer from ${\cal A}$ to ${\cal D}$ takes the form
\begin{equation}
  \begin{pmatrix}{\vec \sigma \cr \vec \psi}\end{pmatrix}=
  \begin{pmatrix}{\cos\alpha & \sin\alpha \cr -\sin\alpha & \cos\alpha }\end{pmatrix}
  \begin{pmatrix}{\vec \rho \cr \vec \phi}\end{pmatrix}\,,
\end{equation}
and the ray transfer from ${\cal D}$ to ${\cal A}'$
\begin{equation}
  \begin{pmatrix}{\vec \rho ' \cr \vec \phi '}\end{pmatrix}=
  \begin{pmatrix}{\cos\alpha '& \sin\alpha '\cr -\sin\alpha '& \cos\alpha '}\end{pmatrix}
  \begin{pmatrix}{\vec \sigma ' \cr \vec \psi '}\end{pmatrix}\,.
\end{equation}
Since $\vec \sigma '=\vec \sigma$ and $\vec \psi '=\vec \psi$, the composition of the two above ray-matrices makes sense and takes the form
\begin{eqnarray}
  \begin{pmatrix}{\vec \rho ' \cr \vec \phi '}\end{pmatrix}&\!\!\!=&\!\!\!
  \begin{pmatrix}{\cos\alpha '& \sin\alpha '\cr -\sin\alpha '& \cos\alpha '}\end{pmatrix}
  \begin{pmatrix}{\cos\alpha & \sin\alpha \cr -\sin\alpha & \cos\alpha }\end{pmatrix}
  \begin{pmatrix}{\vec \rho \cr \vec \phi}\end{pmatrix}\nonumber \\
  &\!\!\!=&\!\!\!\begin{pmatrix}{\cos (\alpha +\alpha ')& \sin (\alpha +\alpha ')\cr -\sin (\alpha +\alpha ') & \cos (\alpha +\alpha ') }\end{pmatrix}
  \begin{pmatrix}{\vec \rho \cr \vec \phi}\end{pmatrix}\,.\label{eq117}
\end{eqnarray}
Equation (\ref{eq117}) expresses the ray transfer from an arbitrary emitter ${\cal A}$ in the object space, to an arbitrary receiver ${\cal A}'$ in the image space.

\bigskip

\noindent {\it (ii) Imaging}

\medskip

The spherical cap ${\cal A}'$ is the coherent image of ${\cal A}$ is $\alpha+\alpha '= 0\;[\pi ]$.

According to Eq.\ (\ref{eq36}) we have (with $q=d+R_A$)
\begin{equation}
\vec \rho=\left[{(R_A+d)(d-R_D+R_{A})\over
    \lambda^2R_A^2 d(R_{D}-d)}\right]^{1/4}\vec r
=\left[{q(q-R_D)\over \lambda^2R_A^2d(R_D-d)}\right]^{1/4}\vec
r\,.\end{equation}
According to Eq.\ (\ref{eq38}) we have (with $q'=d'+R_{A'}$)
\begin{equation}
\vec \rho '=\left[{(d'+R_{A'})(d'-R_D+R_{A'})\over
    \lambda'^2R_{A'}^2 d'(R_{D}-d')}\right]^{1/4}\vec r'
=\left[{q'(q'-R_D)\over \lambda'^2R_{A'}^2d'(R_D-d')}\right]^{1/4}\vec
r'\,.\end{equation}
We use Eqs.\ (\ref{eq110}) and (\ref{eq111}) and write
\begin{eqnarray}
  {q'(q'-R_D)\over d'(R_D-d')}&=&{q(q-R_D)\over d(R_D-d)}\;{[nR_D+d(n'-n)]^2\over
    [nR_D+q(n'-n)]^2}={q(q-R_D)\over d(R_D-d)}\;{d^2R_D^2\left(\displaystyle{n'\over d'}\right)^2\over q^2R_D^2\left(\displaystyle{n'\over q'}\right)^2}\nonumber \\
 &=& {dq'^2\over qd'^2}\; {q-R_D\over R_D-d}\,.
\end{eqnarray}
We use the radius magnification law between ${\cal A}$ and ${\cal A}'$ 
\begin{equation}
  m_{\rm r}={R_{A'}\over R_A}={n\over n'}\,{d'q'\over dq}\,,\end{equation}
and obtain
\begin{equation}
 {1\over \lambda'^2R_{A'}^2}\; {q'(q'-R_D)\over d'(R_D-d')}={n'^4d^4\over n^4d'^4}\;{1\over \lambda^2R_A^2}\;{q(q-R_D)\over d(R_D-d)}\,.
    \end{equation}
Finally, since $\vec r'=m_{\rm v}\vec r$ ($m_{\rm v}$ is the lateral magnification at vertices between ${\cal A}$ and ${\cal A}'$),  we obtain
\begin{equation}
  \vec \rho '={n'|d |\over n|d'|}m_{\rm v}\vec r =\pm \vec \rho\,.\end{equation}

According to Eq.\ (\ref{eq88}), we have $n'm_{\rm v}\vec \Phi '= n\vec \Phi$, and $\vec \Phi$ and $\vec \Phi '$ are colinear. Then $\vec \phi$ and $\vec \phi '$ are colinear. Since $\vec r'=m_{\rm v}\vec r$, we also have
$ n'\vec r'\vec\cdot\vec \Phi '=n\,\vec r\vec\cdot\vec \Phi$, and from  Eqs.\ (\ref{eq37s}) and (\ref{eq38s}) we deduce $n'\lambda '\vec \rho '\vec\cdot \vec \phi '=n\lambda \vec \rho \vec\cdot \vec \phi $. From $n'\lambda '=n\lambda$ and from $\vec\rho '=\pm \vec \rho$, and since $\vec \phi'$ and $\vec\phi$ are colinear, we conclude that
$\vec\phi '=\pm\vec \phi$. (More precisely, we have $\vec\phi '=\vec \phi$, if $\vec \rho '=\vec \rho$, and $\vec\phi '=-\vec \phi$, if $\vec \rho '=-\vec \rho$.)

%********************************************************

%********************************************************
\end{document}